\documentclass{nrc1}
\usepackage{epsfig,paralist,amssymb,cite}

\begin{document}

\title{The Pioneer anomaly: seeking an explanation in newly recovered data}
\author{Viktor T Toth}
\address[label1]{Ottawa, ON K1N 9H5, CANADA}
\author{Slava G Turyshev}
\address[label2]{Jet Propulsion Laboratory, California Institute of Technology, Pasadena, CA 91109, USA}
\shortauthor{Viktor T Toth and Slava G Turyshev}

\maketitle

\begin{abstract}

The Pioneer 10 and 11 spacecraft yielded very accurate navigation in deep space that was, however, limited by a small, anomalous frequency drift of their carrier signals received by the radio-tracking stations of the NASA Deep Space Network (DSN).  This drift signifies a discrepancy between the frequency of the Doppler signals observed by the DSN and that frequency modeled using modern-day deep space navigational codes. This discrepancy, evident in the data for both spacecraft, was interpreted as an approximately constant acceleration. This acceleration has become known as the Pioneer anomaly. The origin of this anomaly is yet unknown. Recent efforts to explain the effect included a search for independent confirmation, analyses of conventional mechanisms, even ideas rooted in new physics, and proposals for a dedicated mission.  We assert that to determine the origin of the Pioneer anomaly, and especially before any discussion of new physics and (or) a dedicated mission can take place, one must analyze the entire set of radiometric Doppler data received from Pioneer 10 and 11. In this paper, we report on our efforts to recover and utilize the complete set of radio Doppler and telemetry records of the Pioneer 10 and 11 spacecraft.  The collection of radio Doppler data for both missions is now complete; we are ready to begin its evaluation. We also make progress utilizing the recently recovered Pioneer telemetry data in the development of a complete engineering profile of the spacecraft. We present a strategy for studying the effect of on-board-generated small forces with this newly recovered information on the performance of thermal, electrical, power, communication, and other subsystems, in conjunction with the analysis of the entire set of the Pioneer Doppler data. We report on the preparations for the upcoming analysis of the newly recovered data with the ultimate goal of determining the origin of the Pioneer anomaly. Finally, we discuss implications of our on-going research of the Pioneer anomaly for other missions, most notably for New Horizons, NASA's recently launched mission to Pluto.
\end{abstract}


\section{Introduction}
\subsection{The Pioneer missions}

The first spacecraft to leave the inner solar system \cite{Dyal-74,HALL1974,JPL2002}, Pioneers 10 and 11 were designed to conduct an exploration of the interplanetary medium beyond the orbit of Mars and perform close-up observations of the planet Jupiter during the 1972--73 Jovian opportunities.

\begin{figure}[h!]
\centering
\includegraphics[width=0.8\linewidth]{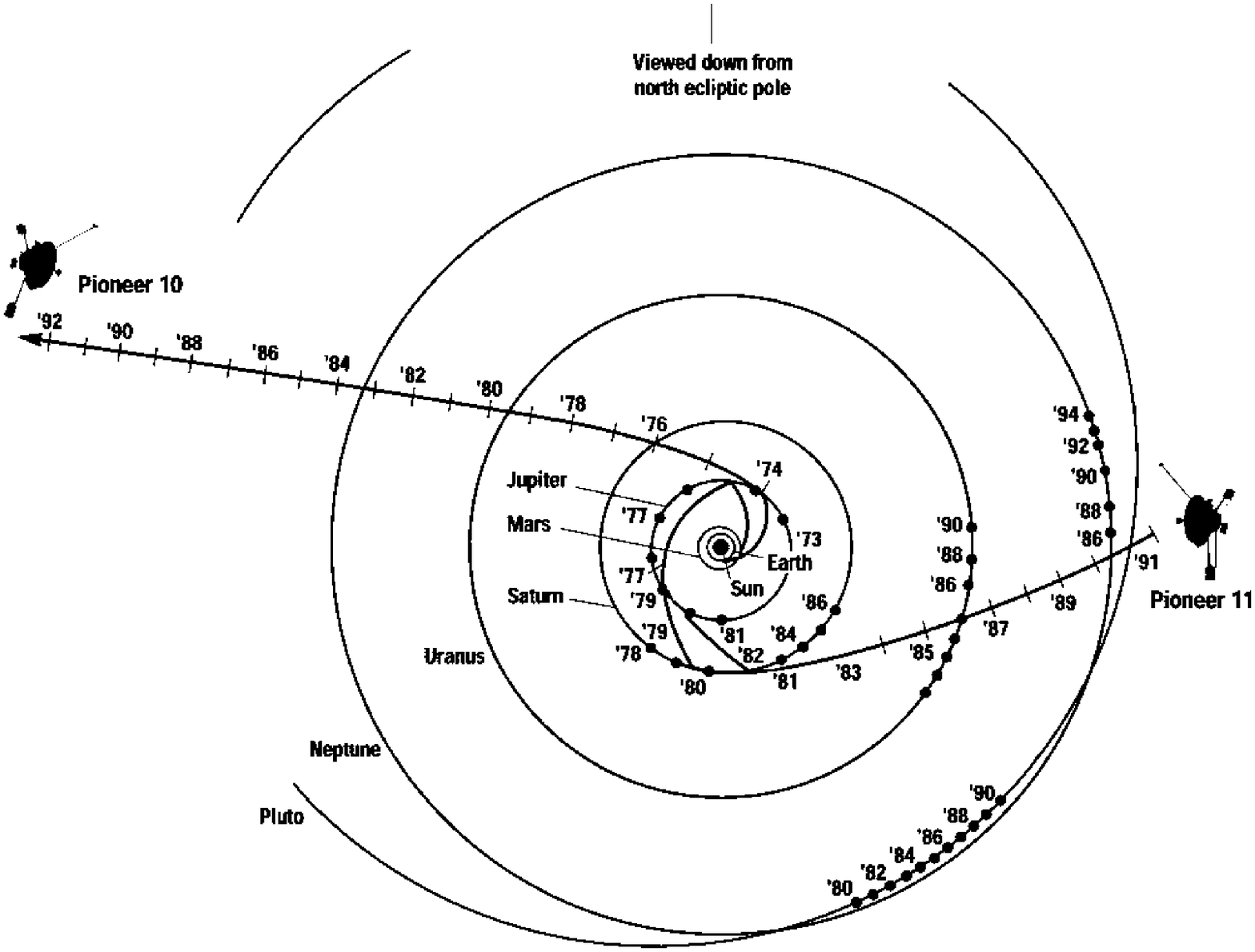}
\caption{Trajectories of the Pioneer spacecraft through the solar system.}
\label{fig:trajectory}
\end{figure}

The spacecraft were launched in March 1972 (Pioneer 10) and April 1973 (Pioneer 11) on top of identical three-stage Atlas--Centaur launch vehicles. After passing through the asteroid belt, Pioneer 10 reached Jupiter in December 1973. The trajectory of its sister craft, Pioneer 11, in addition to visiting Jupiter later that year, also included an encounter with Saturn in 1979 (see Fig.~\ref{fig:trajectory}).

After the planetary encounters and successful completion of their primary missions, both Pioneers continued to explore the outer solar system. Due to their excellent health and navigational capabilities, the Pioneers were used to search for trans-Neptunian objects and to establish limits on the presence of low-frequency gravitational radiation \cite{JPL2002}.

Eventually, Pioneer 10 became the first man-made object to leave the solar system, with its official mission ending in March 1997. Since then, NASA's Deep Space Network (DSN) made occasional contact with the spacecraft.  The last successful communication from Pioneer 10 was received by the DSN on 27 April 2002.\footnote{The last attempt to contact Pioneer 10 took place on the 34$^\mathrm{th}$ anniversary of its launch, on 3--5 March 2006 \cite{new-data-2005}. At that time, the spacecraft was 90.08~AU from the Sun, moving at 12.08~km/s. The round-trip light time (i.e., time needed for a DSN radio signal to reach Pioneer 10 and return back to the Earth) was approximately 24~h~56~m, so the same antenna, DSS-14 at Goldstone, CA, was used for the track. Unfortunately, no signal was received. Given the age of the spacecraft's power source, it is safe to assume that there is no longer sufficient electrical power on board to operate the transmitter.}
Pioneer 11 sent its last coherent Doppler data in October 1990; the last scientific observations were returned by Pioneer 11 in September 1995.

\begin{figure}[h!]
\centering
\includegraphics[width=\linewidth]{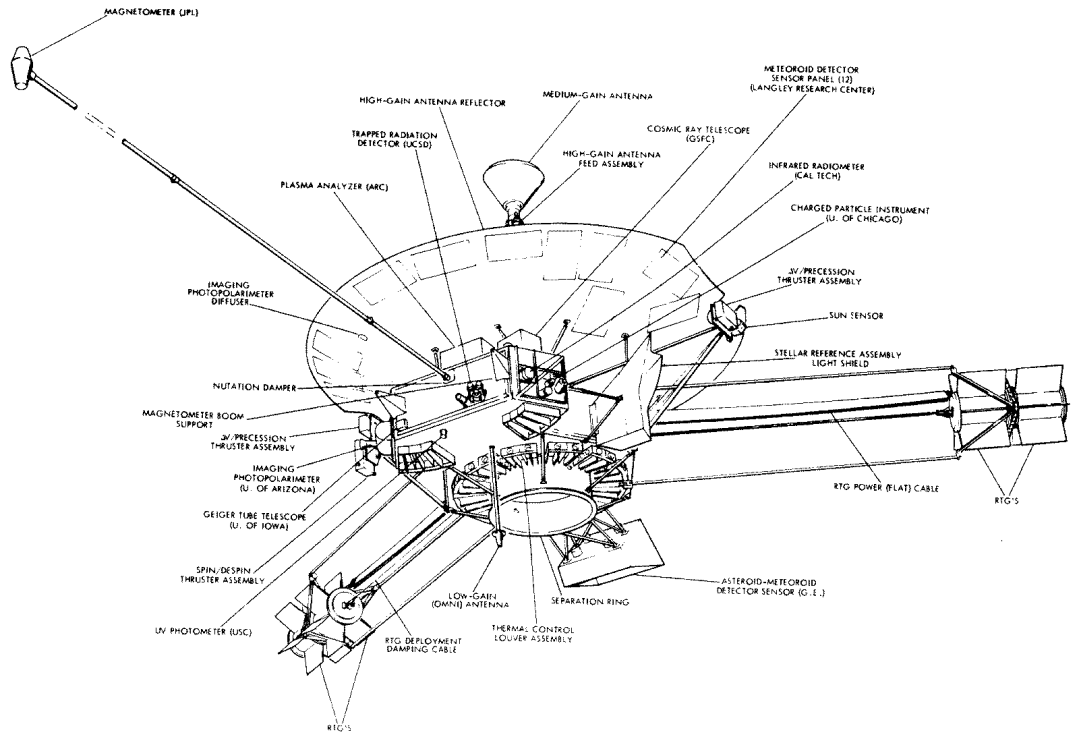}
\caption{A drawing of the Pioneer spacecraft.}
\label{fig:pioneer}
\end{figure}

\subsection{The Pioneer spacecraft}

As evident from Fig.~\ref{fig:pioneer}, the appearance of the spin-stabilized Pioneer spacecraft is dominated by the 2.74~m high gain antenna (HGA). The spacecraft body, located behind the HGA, consists of a larger, regular hexagonal compartment housing the propellant tank and spacecraft electronics; an adjacent, smaller compartment housed science instruments. The spacecraft body is covered by multilayer thermal insulating blankets, except for a louver system located on the side opposite the HGA, which was activated by bimetallic springs to expel excess heat from the spacecraft.

Each spacecraft was powered by four radioisotope thermoelectric generators (RTGs) mounted in pairs at the end of two booms, approximately three meters in length, extended from two sides of the spacecraft body at an angle of 120$^\circ$. A third boom, approximately 6~m long, held a magnetometer.

The total (design) mass of the spacecraft was approximately 250~kg at launch, of which $\sim$27~kg was propellant \cite{PC202} (actual launch masses were 256.51 kg for Pioneer 10, 258.32 kg for Pioneer 11, including propellant and pressurant mass of 25.76 kg and 27.67~kg, respectively \cite{PFG100151}).

For the purposes of attitude control, the spacecraft were designed to spin at the nominal rate of 4.8~rpm. Six small monopropellant (hydrazine) thrusters, mounted in three thruster cluster assemblies, were used for spin correction, attitude control, and trajectory correction maneuvers (see Fig.~\ref{fig:pioneer}).

\subsection{The Pioneer anomaly}

Unique characteristics of Pioneers 10 and 11 made it possible to perform celestial mechanics experiments to an unprecedented (and still unsurpassed) precision \cite{JPL2002,JPL1998,moriond}. The Pioneer spacecraft were spin stabilized, which meant that very few maneuvers were required to maintain spacecraft attitude. The RTG power sources were located at the end of long booms, reducing thermal effects on the spacecraft and improving their stability. As a result, Pioneers 10 and 11 both had an acceleration sensitivity of $\sim10^{-10}$~m/s$^2$.

The orbits of Pioneers 10 and 11 were reconstructed based primarily on radio-metric (Doppler) tracking data. The reconstruction between heliocentric distances of 20$--$70 AU yielded a persistent small discrepancy between observed and computed values \cite{JPL2002,JPL1998,moriond}. This anomalous shift in the Doppler frequency, changing at an approximately constant rate of $\dot{f}_P\sim6\times10^{-9}$~Hz/s, can be interpreted as an unmodeled acceleration towards the inner solar system.

The relationship between the Doppler shift $\dot{f}_P$ and acceleration $a_P$ can be expressed to first order in $v/c$ as

\begin{equation}
a_P=c\frac{\dot{f}_P}{f_0},
\end{equation}
where $f_0\simeq 2.29$~GHz is the reference transmission frequency.

After accounting for known systematic effects \cite{JPL2002}, the unmodeled change in the Doppler residual for Pioneer 10 and 11 is equivalent to an approximately sunward constant acceleration of
\begin{equation}
a_P=(8.74\pm1.33)\times 10^{-10}~~{\rm m/s}^2.
\label{eq:aP}
\end{equation}
The nature of this acceleration remains unexplained; this signal has become known as the Pioneer anomaly.

The magnitude of the anomalous acceleration, measured between heliocentric distances of 40$--$70 AU, remains approximately constant within the 3~dB gain bandwidth of the HGA.

\section{Efforts to explain the anomaly}

There were numerous attempts in recent years to provide an explanation for the anomalous acceleration of Pioneers 10 and 11. These can be broadly categorized as either invoking conventional mechanisms or utilizing principles of ``new physics''.

\subsection{Conventional physics mechanisms}

Initial efforts to explain the Pioneer anomaly focused on the possibility of on-board systematic forces. While these cannot be conclusively excluded \cite{JPL2002,pio-standard}, the evidence to date does not support these mechanisms: the magnitude of the anomaly exceeds the acceleration that these mechanisms would likely produce, and the temporal evolution of the anomaly differs from that which one would expect, for instance, if the anomaly were due to thermal radiation of a decaying nuclear power source.

Conventional mechanisms external to the spacecraft were also considered. First among these was the possibility that the anomaly may be due to perturbations of the spacecrafts' orbits by as yet unknown objects in the Kuiper belt. Another possibility is that dust in the solar system may exert a drag force, or it may cause a frequency shift, proportional to distance, in the radio signal. These proposals could not produce a model that is consistent with the known properties of the Pioneer anomaly, and may also be in contradiction with the known properties of the orbits of Mars and Jupiter.

In particular, measurements by Ulysses and Galileo place an upper limit on the number of dust grains in the $10^{-18}-10^{-12}$~kg range \cite{dust}. Dust in excess of 0.3 Earth masses in the region beyond Neptune is ruled out by infrared observation. The dust that can be present within these constraints cannot exert a sufficient drag force or induce a frequency shift in the radio signal to explain the Pioneer anomaly.

The value of the Pioneer anomaly happens to be approximately $cH_0$ where $c$ is the speed of light and $H_0$ is the Hubble constant at the present epoch. Attempts were made to exploit this numerical coincidence to provide a cosmological explanation for the Pioneer anomaly, but it has been demonstrated that a cosmological mechanism would produce an effect with the opposite sign \cite{JPL2002}.

\subsection{Possibility for new physics?}

As the search for a conventional explanation for the anomaly appeared unsuccessful, this provided a motivation to seek an explanation in ``new physics''. No such attempt to date produced a clearly viable mechanism for the anomaly.

MOdified Newtonian Dynamics (MOND) \cite{Milgrom,Bekenstein} represents an intriguing possibility, as it provides a possible explanation for the anomalous rotation curves of galaxies by altering the long-range behavior of gravity. Unfortunately, the value for the anomalous Pioneer acceleration produced by MOND is much smaller than the observed acceleration.

Nonsymmetric Gravitational Theory (NGT) \cite{moffat} also provides a plausible mechanism. A variant of this theory, Scalar--Tensor--Vector Gravity (STVG) \cite{moffat05} can account for $a_P$\cite{brownstein-moffat} and also claims to explain galaxy and cluster rotation curves.

The presence of dark matter in the solar system may also offer an explanation. For instance, a dark-matter disk in the outer solar system with a density of $\sim4\times10^{-16}$~kg/m$^3$ could yield the observed value.

Additional theories included proposals to modify gravity with a Yuka\-wa-like or higher order correction; conformal gravity \cite{JPL2002}; and scalar fields \cite{orfeu04}.

\section{Recovery of the extended set of Doppler data}

Existing analyses of the anomalous accelerations of Pioneers 10 and 11 were based on limited data sets that spanned only certain intervals of the spacecrafts' lifetime. The most extensive analysis \cite{JPL2002} utilized 11.5 years of Pioneer 10 Doppler data and only 3.75 years of Pioneer 11 data. Since this study, an effort has been underway to collect, organize, and study the entire Doppler mission record \cite{new-data-2005}. We now have almost 30 years of Pioneer 10 and 20 years of Pioneer 11 Doppler data, most of which have never been used to investigate the Pioneer anomaly.

Doppler data are the measure of the cumulative number of cycles of a spacecraft's carrier frequency received during a specific count interval. The exact precision to which these measurements can be carried out is a function of the received signal strength and station electronics, but it is a small fraction of a cycle.

Doppler tracking data may be ``one-way'' (F1), ``two-way'' (F2), or ``three-way'' (F3). In the case of one-way Doppler, the spacecraft is transmitting at a known frequency; the difference between this value and the frequency observed at the tracking station on Earth is the Doppler observable. Although the Pioneer 10 and 11 spacecraft were capable of producing one-way Doppler data, the stability of the on-board oscillator was not sufficient for precision spacecraft tracking.

In the case of two-way and three-way Doppler measurements, a ground-based station sends a signal of a known frequency to the spacecraft. The spacecraft's radio equipment returns a phase-coherent signal that is then received either by the same ground-based station (two-way Doppler) or by another station (three-way Doppler). Since the frequency of the uplink signal is known to great precision (for more details please consult ref.~\citen{asmar_etal-2004}), this data type can be used for high-precision orbit determination.

Raw Doppler data are generated at DSN tracking stations (see the discussion of the present-day DSN capabilities in ref.~\citen{asmar_etal-2004}). Count intervals for Doppler data can vary from 0.1~s to 10 min, with count times of 10 to 60~s being typical \cite{JPL2002,new-data-2005}.  The Pioneers used S-band ($\sim$2.2 GHz) radio signals to communicate with the DSN. The 1-$\sigma$ accuracy of S-band data is approximately 1~mm/s for a 60~s count interval after being calibrated for transmission media effects. The dominant systematic error that can affect S-band tracking data is ionospheric transmission delays. When the spacecraft is located angularly close to the Sun, with Sun--Earth--spacecraft angles of less than 10$^\circ$, degradation of the data accuracy will occur. S-band data are generally unusable for Sun-Earth-spacecraft angles of less than 5$^\circ$.

The set of all available Pioneer 10 and 11 radiometric Doppler data comprises more than 600 files. These files provide a new opportunity to study the Pioneer anomaly, this time with the entire sets of the radiometric Doppler data received from both Pioneers.

The newly available 1972--2002 data set has some redundancy, but mostly it is a very complete data set assembled for the first time (see Table~\ref{tb:new-doppler-data}). For Pioneer 10, it covers mission events from mid-1973 (including the Jupiter encounter data) until the reception of the last telemetry, on April 27, 2002. For Pioneer 11, coverage is from mid-1974 to late 1994.

\begin{table}[h!]\vskip-10pt
\caption{Radiometric Pioneer 10 and 11 Doppler data used in previous studies and presently available for new analysis.}\vskip-10pt
{\small
\begin{center}
\begin{tabular}{|c|c|c|c|c|}\hline
&\multicolumn{2}{c|}{Data used in the previous analyses}
&\multicolumn{2}{c|}{Currently available data}\\\cline{2-5}
Spacecraft& Time span & Distances, AU & Time span & Distances, AU \\\hline\hline
Pioneer 10 & 03.01.87 -- 22.07.98 & 40.0 -- 70.5 & 08.09.73 -- 27.04.02 & 4.56 -- 80.2\\
Pioneer 11 & 05.01.87 -- 01.10.90 & 22.4 -- 31.7& 10.04.73 -- 11.10.94 & 1.01 -- 41.7\\\hline
\end{tabular}
\label{tb:new-doppler-data}
\end{center}
}\vskip-10pt
\end{table}

In preparation for the upcoming analysis of the complete Pioneer trajectories, the thousands of tracking data passes (between 1973 and 2002 for Pioneer 10 and  between 1973 and 1994 for Pioneer 11) will be analyzed, edited, and processed using a common editing strategy, the same initial conditions, parameter estimation, and noise propagation algorithms. High-rate Doppler data (i.e., 1 record/s) will be used to better determine the Pioneers' spin rates, improve the maneuver data file information, and, in conjunction with newly available spacecraft telemetry (see the discussion in Sect.~\ref{sec:telemetry}), to estimate and (or) calibrate valve gas leaks. Each data pass will be analyzed separately, to remove corrupted data and to determine the spin rate of the spacecraft.

To improve the accuracy of the analysis, we will utilize the latest planetary ephemeris (DE410) and the latest Earth models. This will allow us to better account for the annual and diurnal terms \cite{JPL2002} that were present in the Pioneer 10 and 11 Doppler residuals.

We will also update our analysis programs, to take into account improvements in the models for the Earth's orientation, ephemeris, atmosphere, and ionosphere; new station locations; crustal motion; and atmospheric and ionosphere effects. We will develop algorithms to describe gas leaks and thermal dissipation processes on the Pioneer spacecraft. The engineering, navigational, and telemetry records for the craft will also be updated.  Together with newly available Doppler data, this will allow us to improve the accuracy to which the effects due to these on-board small forces can be evaluated.

\section{Developing the spacecrafts' history from telemetry}
\label{sec:telemetry}

All transmissions of the Pioneer 10 and 11 spacecraft, including all engineering telemetry, were archived \cite{new-data-2005} in the form of files containing Master Data Records (MDRs). Originally, MDRs were scheduled for limited retention. Fortunately, the Pioneers' mission records avoided this fate: with the exception of a few gaps in the data (mainly due to deteriorating media) the entire mission record has been saved and is available for study. Our recent work provides us with the tools necessary to extract telemetry information from these files.

These newly available telemetry readings are important in reconstructing a complete history of the thermal, electrical, and propulsion systems for both spacecraft. This, it is hoped, may in turn lead to a better determination of the spacecrafts' acceleration due to on-board systematic effects.

\subsection{Data organization}

Telemetry formats can be broadly categorized as science formats versus engineering formats. Telemetry words included both analog and digital values. Digital values were used to represent sensor states, switch states, counters, timers, and logic states. Analog readings, from sensors measuring temperatures, voltages, currents and more, were encoded using 6-bit words. This necessarily limited the sensor resolution and introduced a significant amount of quantization noise. Furthermore, the analog-to-digital conversion was not necessarily linear; prior to launch, analog sensors were calibrated using a fifth-order polynomial. Calibration ranges were also established; outside these ranges, the calibration polynomials are known to yield nonsensical results.

With the help of the information contained in these words, it is possible to reconstruct the history of
\begin{itemize}
\item RTG temperatures and power,
\item radio beam power,
\item electrically generated heat inside the spacecraft body,
\item spacecraft temperatures, and
\item the propulsion system.
\end{itemize}

In some cases, the readings are redundant, which provides a means to verify the values, and the methods we use to evaluate them, for consistency. In other cases, it is possible to compare the readings to predictions based on first principles (e.g., the drop in RTG temperatures due to the decay of the radioactive fuel). These exercises can greatly increase our confidence in the data and our evaluation.

Engineering telemetry words are labeled using identifiers in the form of $C_n$, where $n$ is a number indicating the word position in the fixed format telemetry frames.

\subsection{RTG temperatures and power}

\begin{figure*}[h!]
\hskip -6pt
\begin{minipage}[b]{.5\linewidth}
\centering \psfig{file=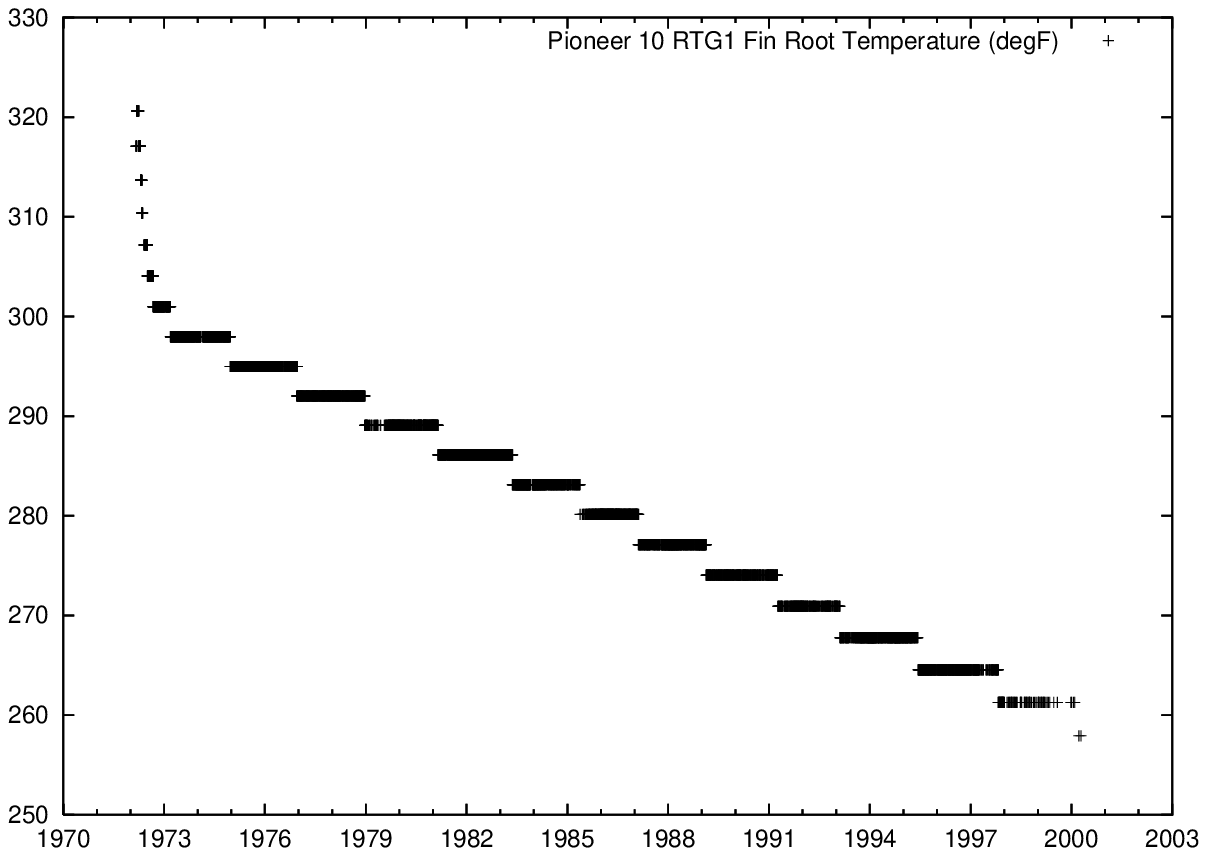, width=\linewidth}
\end{minipage}
\hskip 0.001\linewidth
\begin{minipage}[b]{.5\linewidth}
\centering \psfig{file=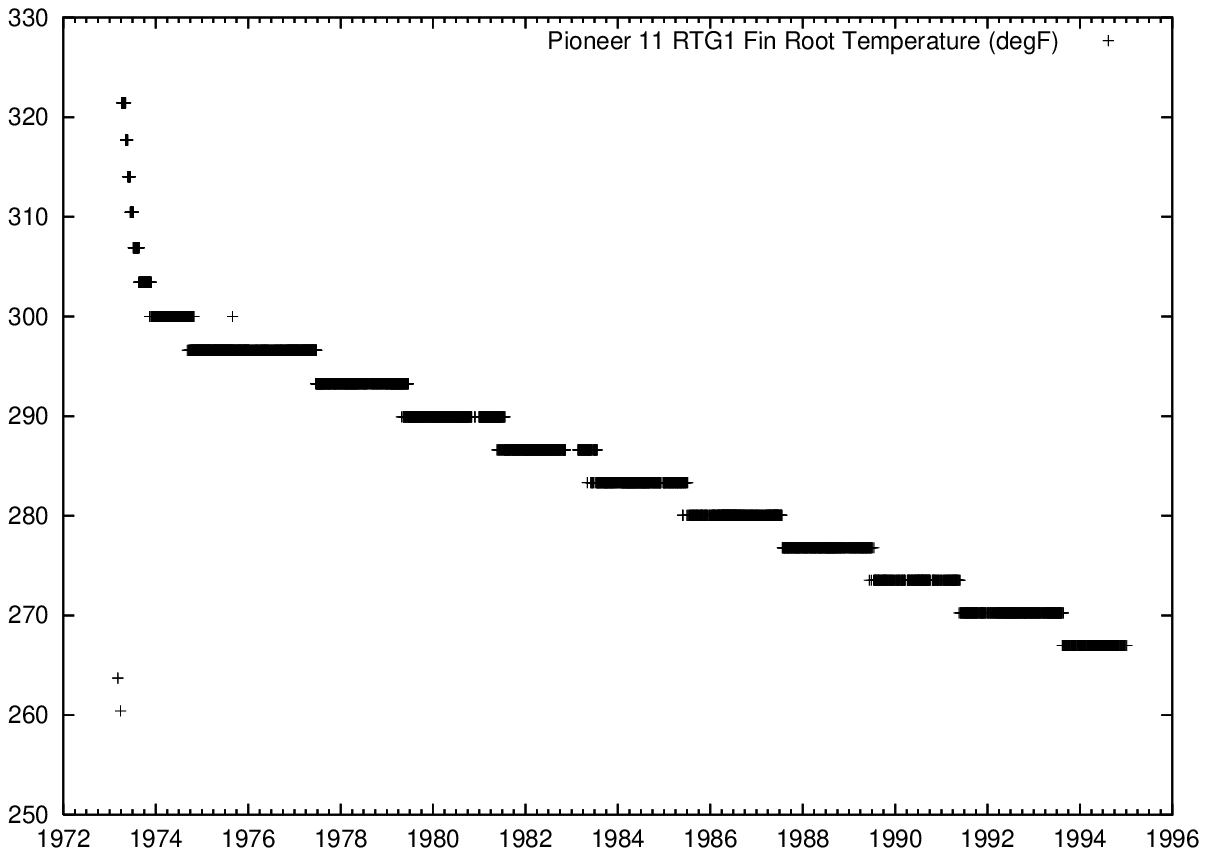, width=\linewidth}
\end{minipage}
\vskip -5pt
\caption{RTG 1 fin root temperatures (telemetry word $C_{201}$; in $^\circ$F) for Pioneer 10 (left) and 11 (right). The behavior of the other RTGs was nearly identical.}
\label{fig:C201}
\end{figure*}

The exterior temperatures of the RTGs were measured by one sensor on each of the four RTGs: the so-called ``fin root temperature'' sensor. Telemetry words $C_{201}$ through $C_{204}$ contain the fin root temperature sensor readings for RTGs 1 through 4, respectively. Figure~\ref{fig:C201} depicts the evolution of the RTG 1 fin root temperature for Pioneers 10 and 11; the other RTGs behaved similarly.

A best fit analysis confirms that the RTG temperature indeed evolves in a manner consistent with the radioactive decay of the nuclear fuel on board.

\begin{figure*}
\hskip -6pt
\begin{minipage}[b]{.5\linewidth}
\centering \psfig{file=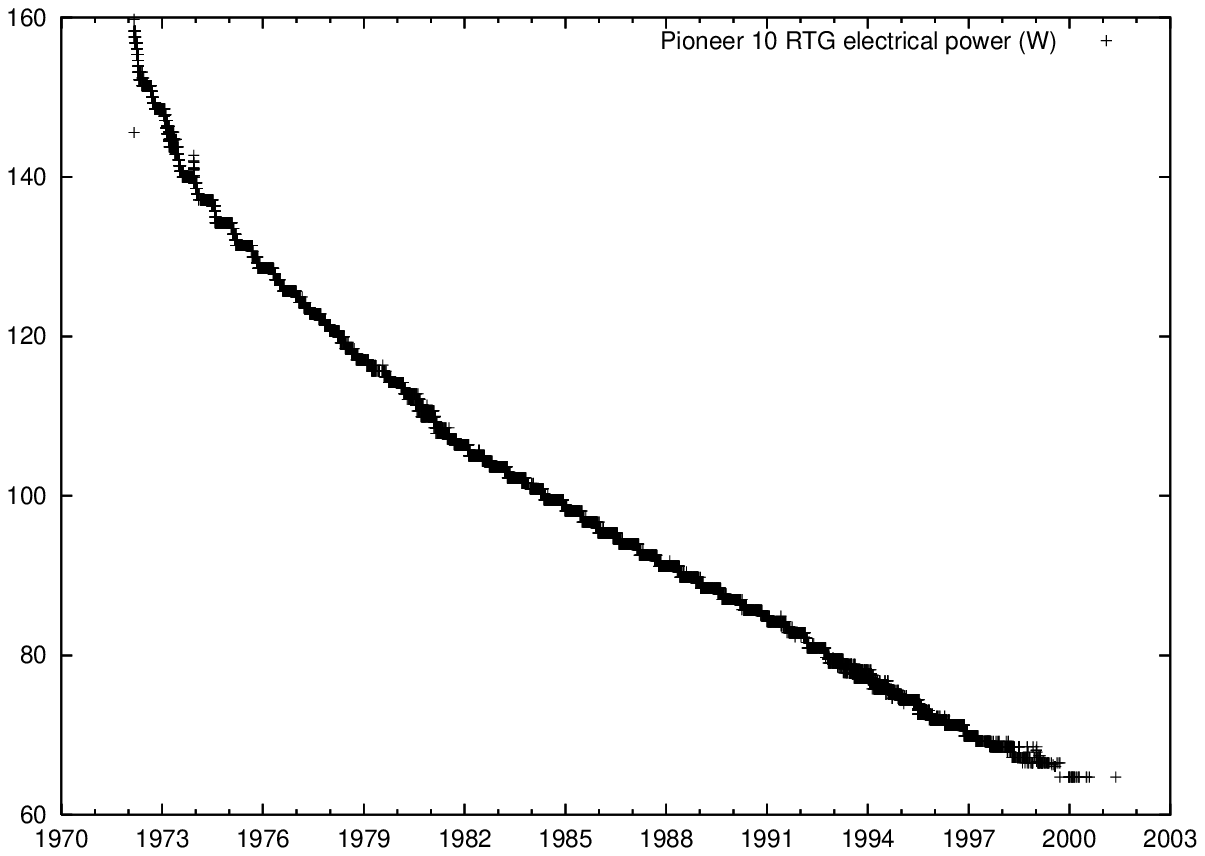, width=\linewidth}
\end{minipage}
\hskip 0.001\linewidth
\begin{minipage}[b]{.5\linewidth}
\centering \psfig{file=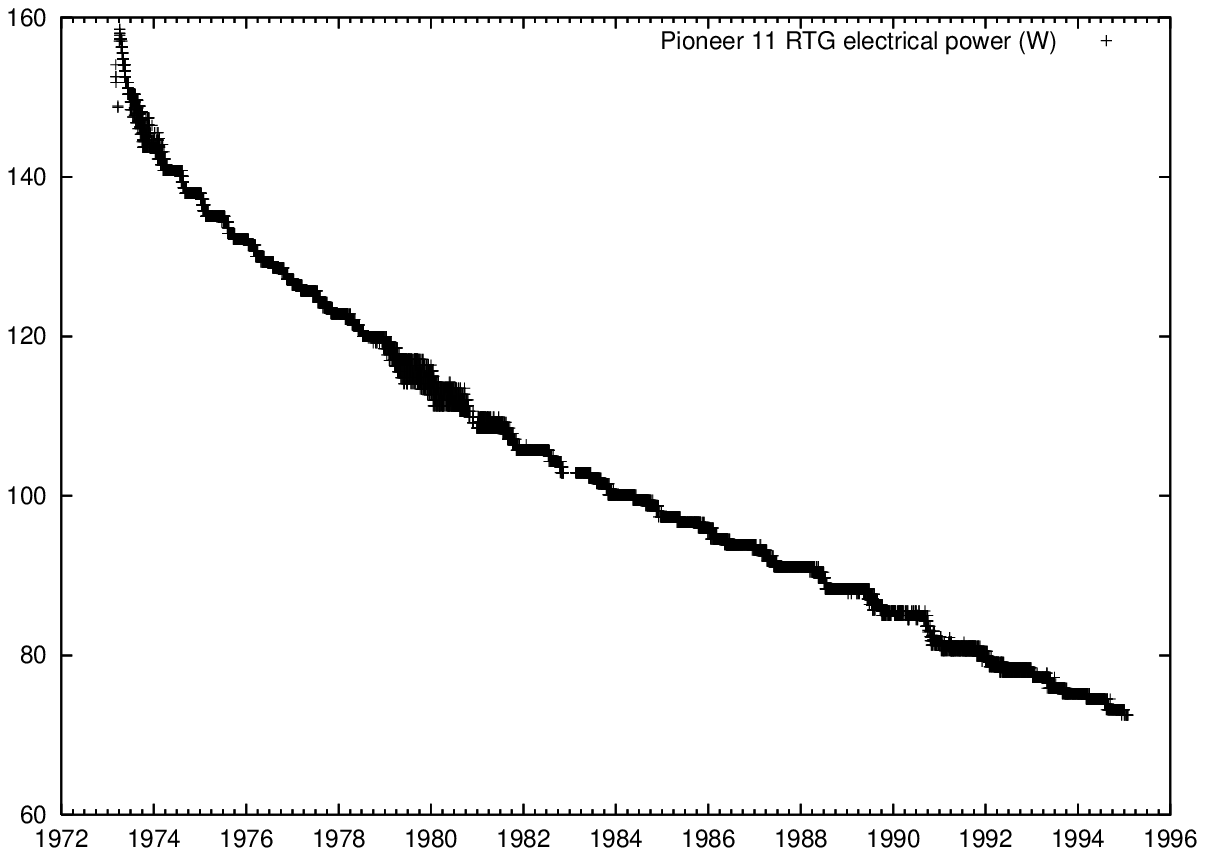, width=\linewidth}
\end{minipage}
\caption{Changes in total RTG electrical output (in W) on board Pioneer 10 (left) and 11 (right), as computed using the missions' on-board telemetry.}
\label{fig:elec}
\vskip -5pt
\end{figure*}

The results for all the other RTGs are similar, confirming that the RTGs were performing thermally in accordance with design expectations.

RTG electrical power can be estimated using two sensor readings per RTG: one measuring RTG current, the other measuring RTG voltage. Currents for RTGs 1 through 4 appear as telemetry words $C_{127}$, $C_{105}$, $C_{114}$, and $C_{123}$, respectively; voltages are in telemetry words $C_{110}$, $C_{125}$, $C_{131}$, and $C_{113}$. Combined, these words yield the total amount of electrical power available on board:
\begin{equation}
P_E=\displaystyle\sum_{n=1}^4U_nI_n=C_{110}C_{127}+C_{125}C_{105}+C_{131}C_{114}+C_{113}C_{123}.
\end{equation}

All this electrical power (Figure~\ref{fig:elec}) is eventually converted to waste heat by the spacecrafts' instruments, with the exception of power radiated away by radio transmitters.

\subsection{Radio transmitter power}

The power of the spacecrafts' traveling wave tube (TWT) microwave transmitters is nominally 8~W. In actuality, however, the transmitter power varied over time, presumably due to the aging of components, changing temperatures inside the spacecraft body, and, near the end of the mission, a drop in the main bus voltage that may have affected transmitter operation.

\begin{figure*}
\hskip -6pt
\begin{minipage}[b]{.5\linewidth}
\centering \psfig{file=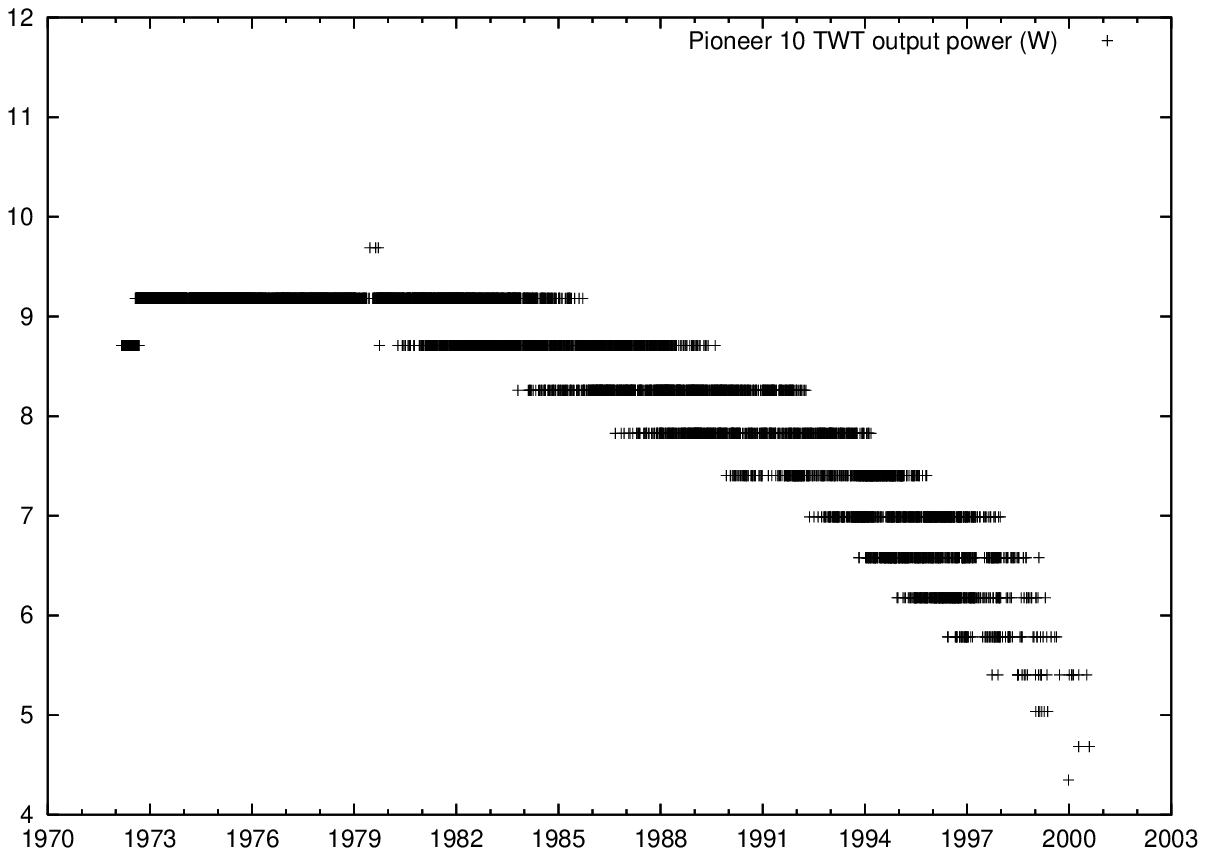, width=\linewidth}
\end{minipage}
\hskip 0.001\linewidth
\begin{minipage}[b]{.5\linewidth}
\centering \psfig{file=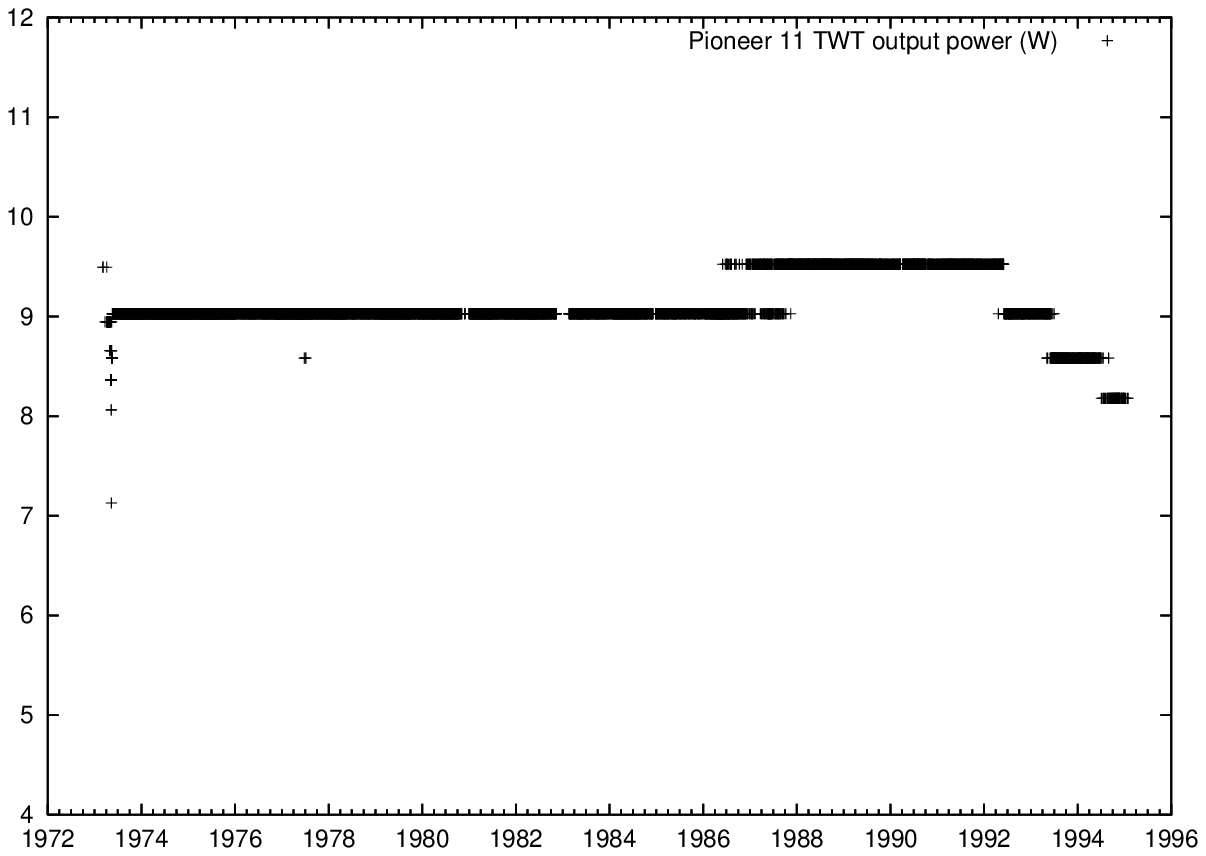, width=\linewidth}
\end{minipage}
\caption{The emitted power (measured in dBm, converted to W) of the traveling wave tube transmitter throughout the mission, as measured by on-board telemetry. Left: Pioneer 10, which used TWT A (telemetry word $C_{231}$). Right: Pioneer 11, initially using TWT A but switching to TWT B (telemetry word $C_{214}$) early in its mission.}
\label{fig:TWT}
\vskip -5pt
\end{figure*}

Telemetry words $C_{231}$ and $C_{214}$ provide direct measurements, in units of dBm, of the transmitter power of the two transmitters on board (only one of which was used at any given time). Figure~\ref{fig:TWT} shows the evolution of transmitter power on board Pioneers 10 and 11 throughout their mission; as can be seen, transmitter power slowly declined, by as much as 3~W towards the end of the mission.

The direct reading of emitted power is not the only measurement characterizing the state of the spacecrafts' transmitters. Another reading, that of the TWT cathode current, appears to confirm that transmitter power varied over time.

\subsection{Automatic Gain Control (AGC)}

\begin{figure*}[h!]
\hskip -6pt
\begin{minipage}[b]{.5\linewidth}
\centering \psfig{file=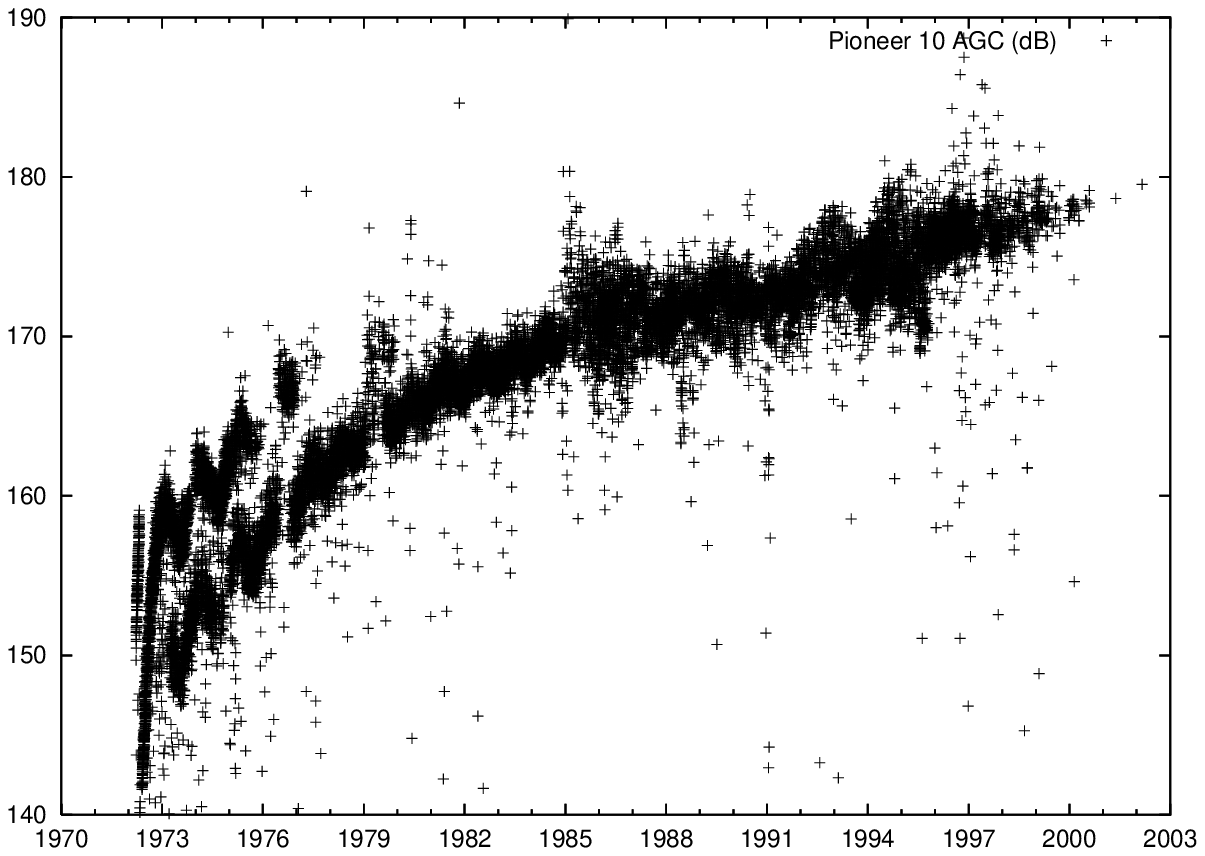, width=\linewidth}
\end{minipage}
\hskip 0.001\linewidth
\begin{minipage}[b]{.5\linewidth}
\centering \psfig{file=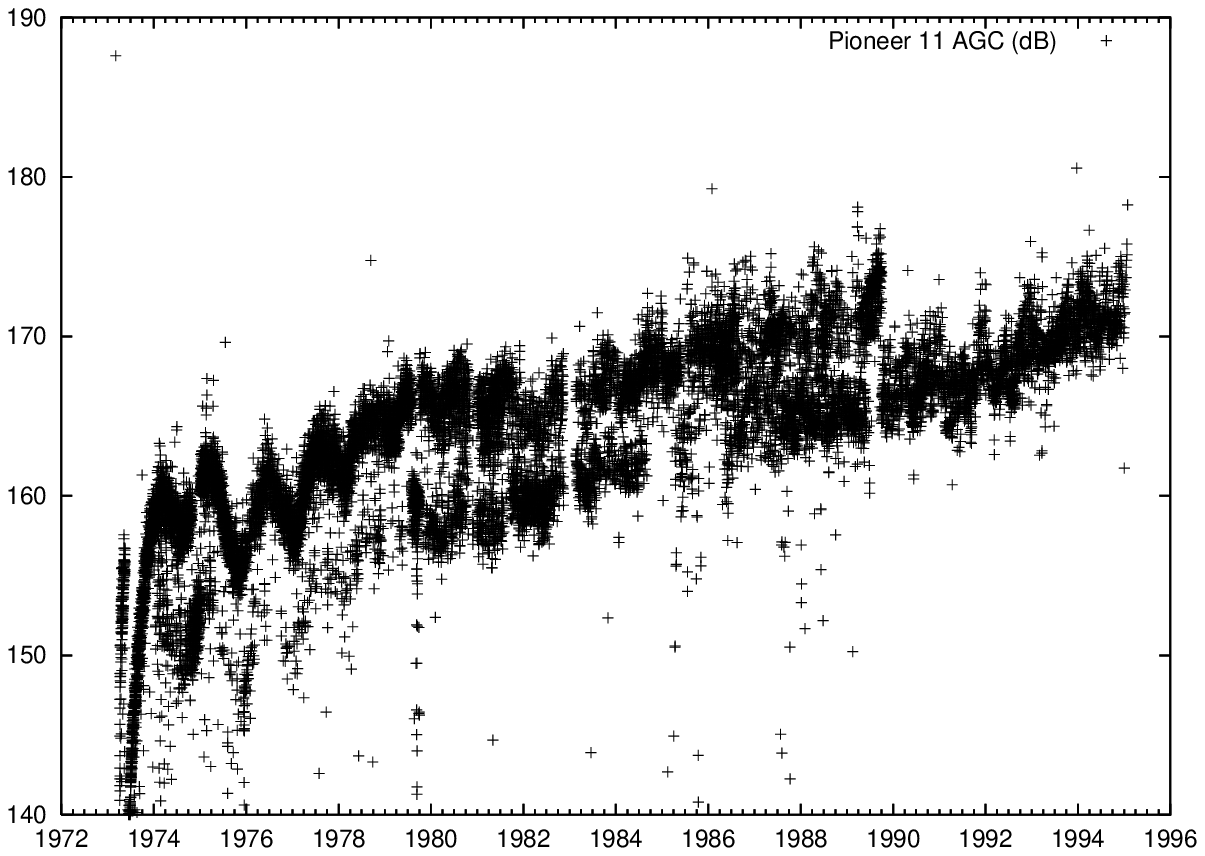, width=\linewidth}
\end{minipage}
\vskip -5pt
\caption{Uncorrected AGC values for Pioneer 10 (left) and 11 (right).}
\label{fig:rawAGC}
\end{figure*}

Closely related to the performance of the spacecrafts' communications systems on board is the strength of the received signal at DSN stations, information that is also available in the telemetry records that were recovered. Specifically, the DSN stations recorded the automatic gain control (AGC) value applied at the time of reception to increase the signal strength to a nominal value. In other words, the actual signal strength was $-1$ times the AGC value, measured in units of dB over the nominal level.

Figure~\ref{fig:rawAGC} shows the AGC for Pioneers 10 and 11. As expected, the signal clearly shows an annual modulation laid over a steady decrease in power, as the spacecraft recedes from the inner solar system while the Earth sometimes moves closer while circling the Sun.

A little more difficult to explain is the ``double'' appearance of the signal especially early in the missions, until one realizes that different DSN antennas were used to receive the signal. The antenna gain of a 34~m DSN antenna is obviously not the same as the gain of a large, 70~m antenna. Fortunately, the Pioneer MDRs also contain DSN station identification information. From historical records, the DSN antenna size and approximate antenna gain can be recovered even for antennas that have long since been decommissioned.

The resolution of the AGC reading in the telemetry records is 0.25~dB (2 digits after the binary point in a fixed point number representation). Our knowledge of the actual antenna gain of various DSN stations can be improved if better historical records are located; alternatively, we may restrict our investigations to data records obtained from a specific DSN antenna. Our hope is that we will be able to reduce the noise that is apparent in Fig.~\ref{fig:rawAGC}, and, despite the limited resolution of the AGC value, we may be able to use this data to improve our understanding of the Pioneers' trajectory and orientation.

\subsection{Electrically generated heat}

\begin{figure}
\centering
\includegraphics[width=\linewidth]{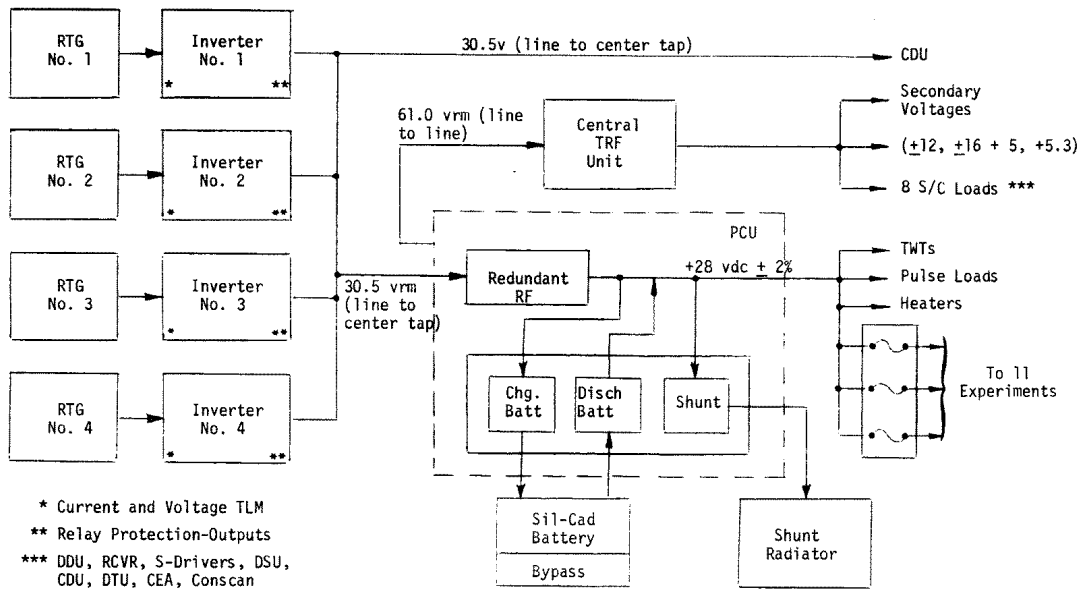}
\caption
{Overview of the Pioneer 10/11 electrical subsystem.}
\label{fig:electrical}
\end{figure}

Whatever remains of electrical energy (see Fig.~\ref{fig:electrical}) after accounting for the power of the transmitted radio beam is converted to heat on board. Some of it will be converted to heat in places external to the spacecraft body.

The Pioneer electrical system is designed to maximize the lifetime of the RTG thermocouples by ensuring that the current draw from the RTGs is always optimal. This means that power supplied by the RTGs may be more than that required for spacecraft operations. Excess electrical energy is absorbed by a shunt circuit that includes an externally mounted radiator plate. Especially early in the mission, when plenty of RTG power was still available, this radiator plate was the most significant component external to the spacecraft body that radiated heat. From design documentation \cite{PC202}, it is possible to deduce that the external shunt radiator is a $5.25~\mathrm{\Omega}$ resistor. We have a telemetry word, $C_{122}$, that tells us the shunt circuit current, from which the amount of power dissipated by the external radiator can be computed.

Other externally mounted components that consume electrical power are the Plasma Analyzer ($P_{\mathrm{PA}}=4.2~\mathrm{W}$, telemetry word $C_{108}$ bit 2), the Cosmic Ray Telescope ($P_{\mathrm{CRT}}=2.2~\mathrm{W}$, telemetry word $C_{108}$, bit 6), and the Asteroid/Meteoroid Detector ($P_{\mathrm{AMD}}=2~\mathrm{W}$, telemetry word $C_{124}$, bit 5). Though these instruments' exact power consumption is not telemetered, we know their average power consumption, and the telemetry bits tell us when these instruments were powered.

Two additional external loads are the battery heater and the propellant line heaters. These represent a load of $P_{\mathrm{LH}}=P_{\mathrm{BH}}=2~\mathrm{W}$ (nominal) each. The power state of these loads is not telemetered. According to mission logs, the battery heater was commanded off on both spacecraft on 12 May 1993 \cite{PNAGENDA93}.

Yet a further external load is the set of cables connecting the RTGs to the inverters.\footnote{It is stated in some of the design documentation that the RTG voltage sensors measure voltage at the inverters, i.e., after cable losses are taken into account. An analysis of the data, however, suggests that the sensors may have been calibrated to measure actual RTG voltages. In any case, the difference is small, as cable losses are small.} The resistance of these cables is known \cite{PC202}: it is $0.017~\mathrm{\Omega}$ for the inner RTGs (RTG 3 and 4), and $0.021~\mathrm{\Omega}$ for the outer RTGs (RTG 1 and 2). Using the RTG current readings it is possible to accurately determine the amount of power dissipated by these cables in the form of heat:
\begin{equation}
P_{\mathrm{cable}}=\displaystyle\sum_{n=1}^4R_nI_n^2=0.017\left(C_{114}^2+C_{123}^2\right)+0.021\left(C_{127}^2+C_{105}^2\right).
\end{equation}

After accounting for all these external loads, whatever remains of the available electrical power on board is converted to heat inside the spacecraft. So long as the body of the spacecraft is in equilibrium with its surroundings, heat dissipated through its walls has to be equal to the heat generated inside:
\begin{equation}
P_{\mathrm{body}}=P_E-P_{\mathrm{cable}}-P_{\mathrm{PA}}-P_{\mathrm{CRT}}-P_{\mathrm{AMD}}-P_{\mathrm{LH}}-P_{\mathrm{BH}},
\end{equation}
with all the terms defined above.

\subsection{Compartment temperatures and thermal radiation}

The infrared radiometric properties of the exterior of the spacecraft body are known. In particular, we have two types of exterior surfaces: those covered with multilayer insulating thermal blankets, and those covered with the spacecrafts' passive thermal control system.

\begin{figure}[h!]
\centering
\includegraphics[width=0.6\linewidth]{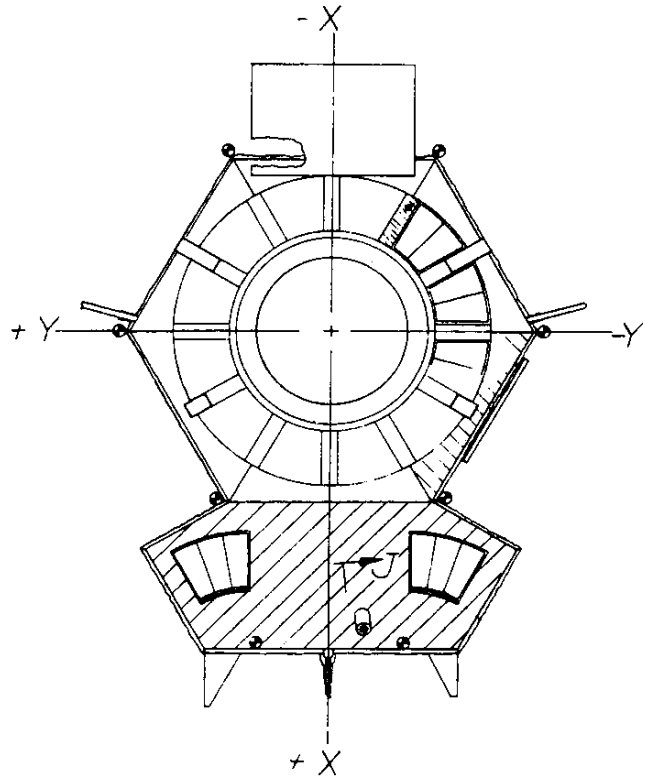}
\caption{Bottom view of the Pioneer 10 and 11 spacecraft, showing the louver system (from \cite{TCSDR3}). A set of 12 2-blade louver assemblies cover the main compartment in a circular pattern; an additional two 3-blade assemblies cover the compartment that contains science instruments.}
\label{fig:louvers}
\end{figure}

The passive thermal control system consisted of a series of spring-activated louvers (see Fig.~\ref{fig:louvers}). The springs were bimetallic, and thermally (radiatively) coupled to the electronics platform beneath the louvers. The louver blades were highly reflective in the infrared. The assembly was designed so that the louvers fully open when temperatures reach $30^{\circ}$C, and fully close when temperatures drop below $5^{\circ}$C \cite{PC202}.

The effective emissivity of the thermal blankets used on the Pioneers is $\epsilon_{\mathrm{sides}}=0.085\pm 0.015$ \cite{SJ1972, LKS2003}. The total exterior area of the spacecraft body is $A_{\mathrm{walls}}=4.92~\mathrm{m}^2$ \cite{PC202}.

The front side of the spacecraft body that faces the HGA has an area of $A_{\mathrm{front}}=1.53~\mathrm{m}^2$, and its effective emissivity, accounting for the fact that most thermal radiation this side emits is reflected by the back of the HGA, can be computed as $\epsilon_{\mathrm{front}}=0.0013$.

The area covered by louver blades is $A_{\mathrm{louver}}=0.29~\mathrm{m}^2$ \cite{PC202}; the effective emissivity of closed louvers is $\epsilon_{\mathrm{louver}}=0.04$ \cite{PC202}.

The area that remains, consisting of the sides of the spacecraft and the portion of the rear not covered by louvers is $A_{\mathrm{sides}}=A_{\mathrm{walls}}-A_{\mathrm{front}}-A_{\mathrm{louver}}$.

Using these numbers, we can compute the amount of electrically generated heat radiated through the (closed) louver system as a ratio of total electrical heat generated inside the spacecraft body:
\begin{equation}
P_{\mathrm{louver}}=\frac{\epsilon_{\mathrm{louver}}A_{\mathrm{louver}}P_{\mathrm{body}}}{\epsilon_{\mathrm{louver}}A_{\mathrm{louver}}+\epsilon_{\mathrm{sides}}A_{\mathrm{sides}}+\epsilon_{\mathrm{front}}A_{\mathrm{front}}}.
\label{eq:eheat}
\end{equation}

This result is a function of the electrical power generated inside the spacecraft body. However, we also have in our possession thermal vacuum chamber test results of the Pioneer louver system. These results characterize louver thermal emissions as a function of the temperature of the electronics platform beneath the louvers, with separate tests performed for the 2-blade and 3-blade louver assemblies. To utilize these results, we turn our attention to telemetry words representing electronics platform temperatures in Pioneer telemetry.

\begin{figure}[h!]
\centering
\includegraphics[width=0.6\linewidth]{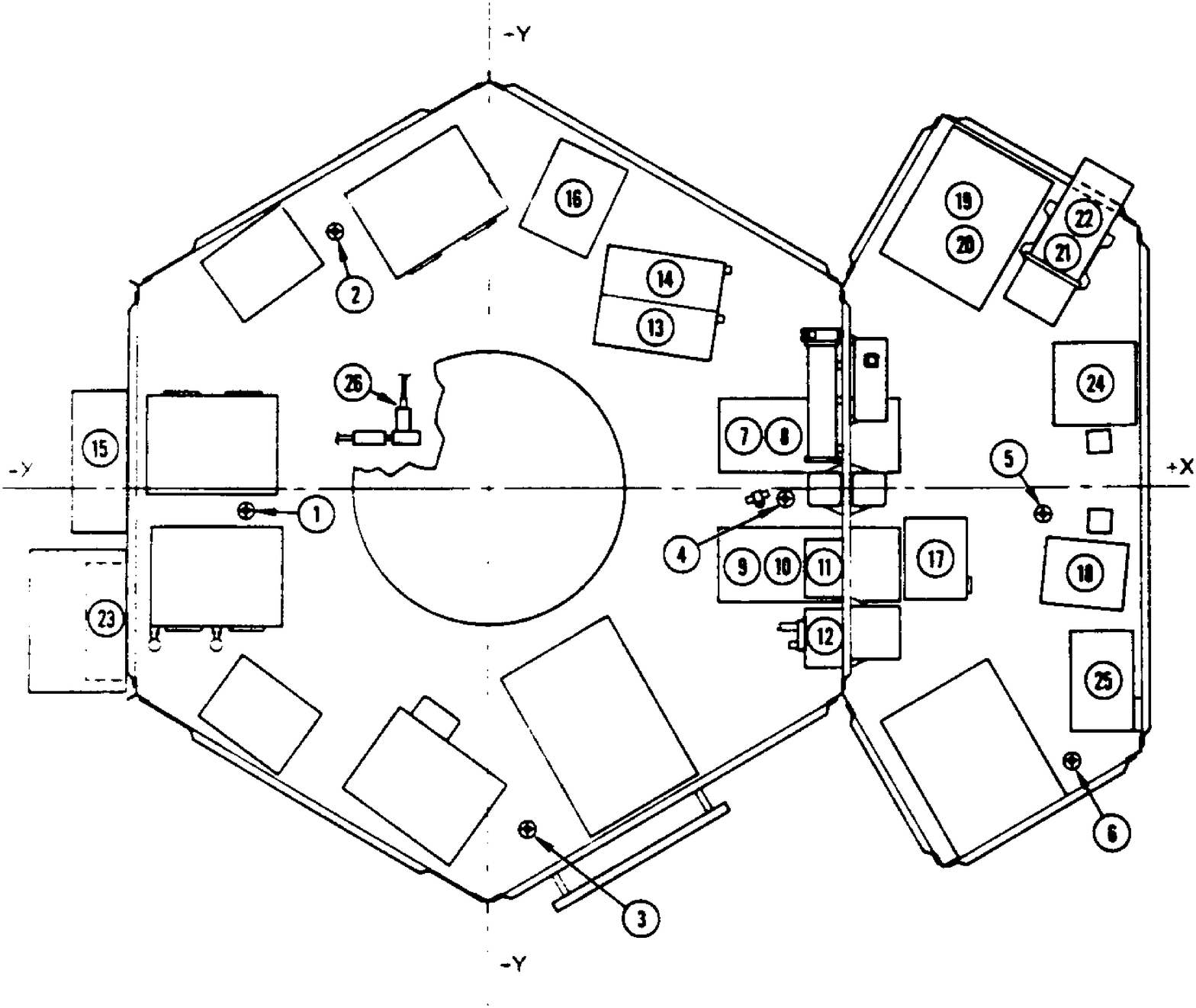}
\caption{Location of thermal sensors in the instrument compartment of the Pioneer 10 and 11 spacecraft \cite{PC202}. Platform temperature sensors are mounted at locations 1 through 6.}
\label{fig:tempsens}
\end{figure}

There are six platform temperature sensors (Fig.~\ref{fig:tempsens}) inside the spacecraft body: four are situated inside the main compartment, while an additional two sensors are located in the science instrument compartment. The main compartment has a total of 12 2-blade louver blade assemblies; the science compartment has 2 3-blade assemblies.

The thermal vacuum chamber tests, documented in ref.~\citen{TCSDR3}, provide values for emitted thermal power per louver assembly as a function of the temperature of the electronics platform behind the louver. Averaging the four platform temperatures in the main compartment, and the two platform temperatures in the science compartment, we can estimate the amount of thermal power leaving the spacecraft body through the louvers, as a function of platform temperatures. This estimate is independent of that developed in (\ref{eq:eheat}), which was based on available electrical heat and the geometry of the spacecraft body.

\begin{figure}[h!]
\begin{minipage}[b]{.5\linewidth}
\centering
\includegraphics[width=\linewidth]{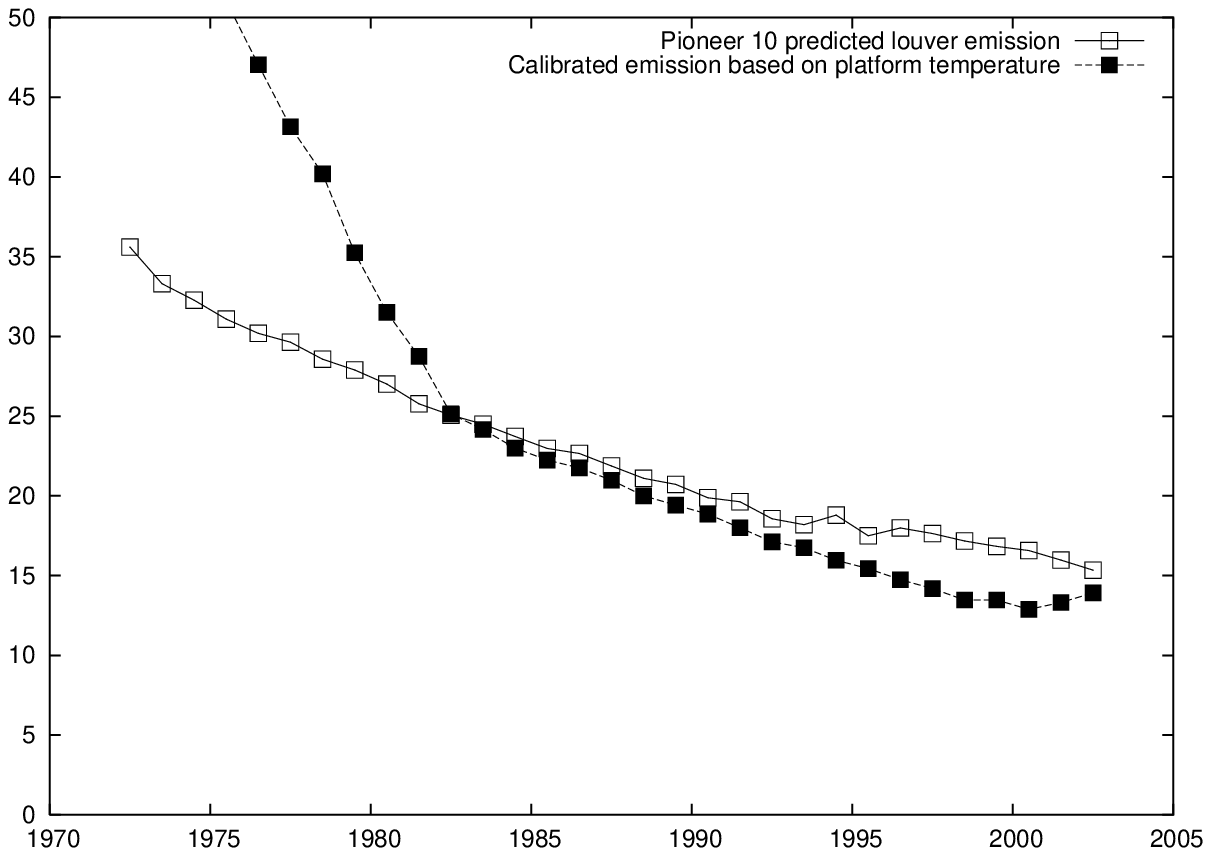}
\end{minipage}
\hskip .001\linewidth
\begin{minipage}[b]{.5\linewidth}
\centering
\includegraphics[width=\linewidth]{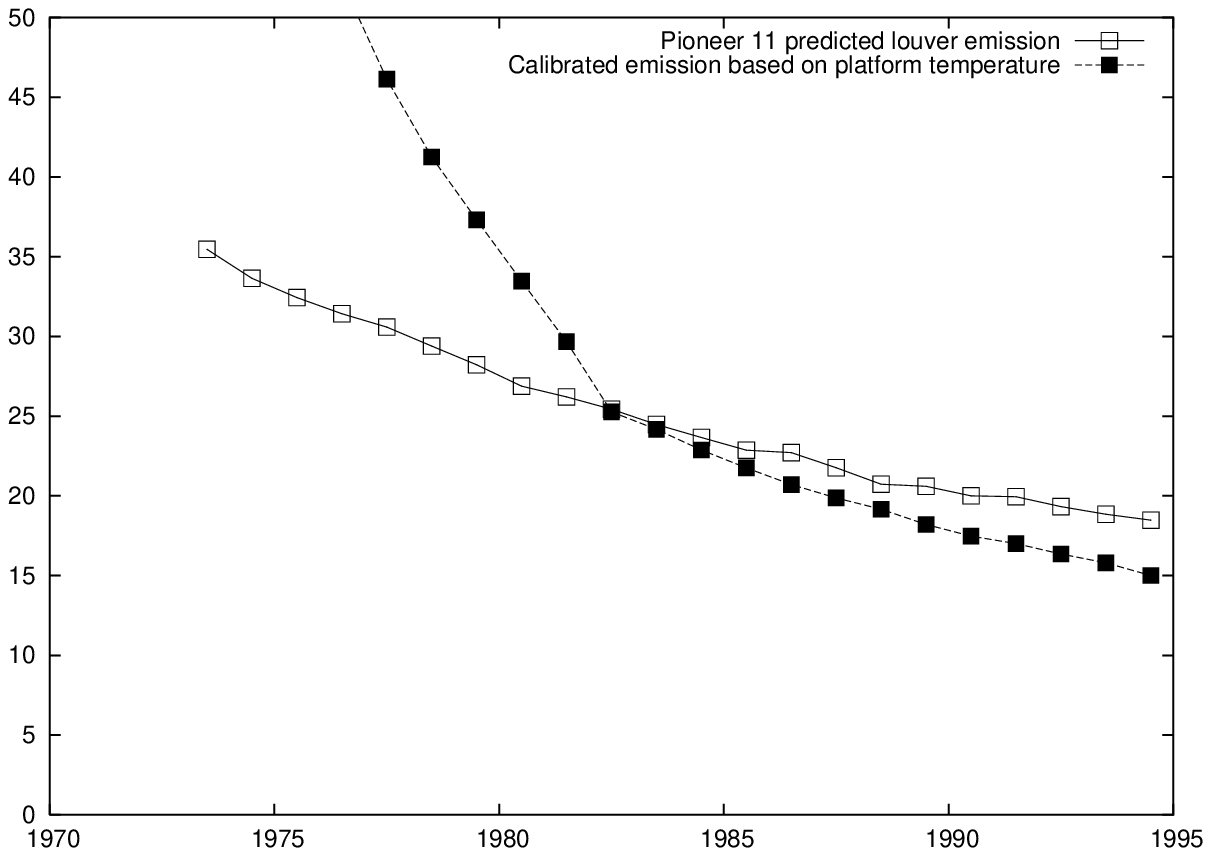}
\end{minipage}
\caption{Pioneer 10 (left) and 11 (right) louver thermal emission estimates (in W). Values predicted on the basis of available electrical power and values based on platform temperatures and louver system calibration data are shown. Predictions based on electrical power do not take into account the increased effective emissivity of the louvers when they open; once temperatures fell below 40$^\circ$F and the louvers were fully closed, the two results are in good agreement.}
\label{fig:louver}
\end{figure}

Figure~\ref{fig:louver} shows the results of this analysis. As can be seen, with the exception of early data from a time when the louvers were not yet fully closed (thus, their effective emissivity was higher than modeled here), the agreement between the two estimates is quite good. This allows us to conclude that the model used to estimate the amount of thermal power generated inside the spacecraft body and the thermal power emitted through the spacecraft walls and the louver system may be useful in further analysis.

\subsection{Propulsion}

The analysis of Anderson et al \cite{JPL2002} demonstrates that three distinct intervals exist in Pioneer 10's acceleration and spin history during the 11.5 years that this analysis examined. Now that we are in possession of the full telemetry record, one question immediately arises: are these intervals visible in telemetry?
\begin{figure}[h!]
\centering
\includegraphics[width=\linewidth]{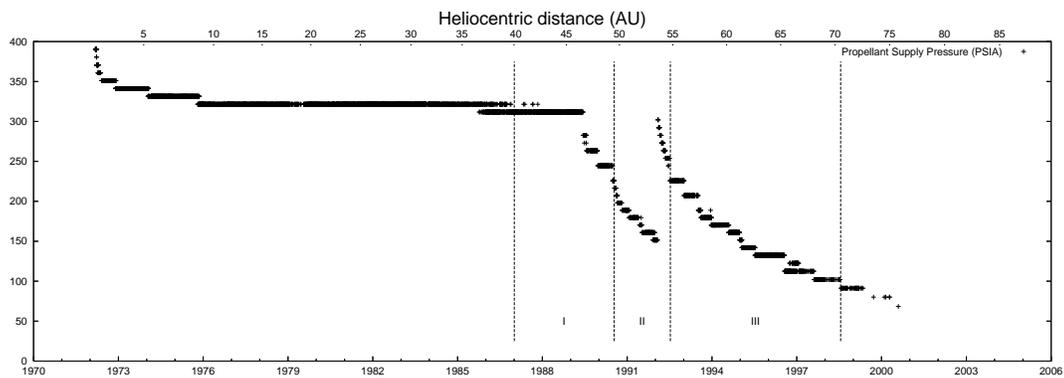}
\caption{Propulsion tank pressure (in psia) on board Pioneer 10. The three intervals studied in ref.~\citen{JPL2002} are marked by roman numerals and separated by vertical lines.}
\label{fig:c210}
\end{figure}

The answer is affirmative, as demonstrated by Fig.~\ref{fig:c210}. This figure shows the propellant tank pressure (telemetry word $C_{210}$). Changes in the behavior of this reading, approximately coinciding with interval II discussed in \cite{JPL2002}, are clearly visible. No explanation is presently known for these changes\footnote{The most puzzling is the sudden doubling in the propellant supply pressure at the beginning of 1993, shown in interval II of Fig.~\ref{fig:c210}.  Most likely this apparent jump is the result of a sensor malfunction, but it may also be related to the propulsion maneuver conducted on 5 January 1993. A direct mechanism for this connection needs further investigation. }; however, the fact that they coincide with the interval boundaries is strongly indicative of the possibility that the two are, in fact, related.

\section{A strategy to find the origin of the Pioneer anomaly}

Our continuing effort to process and analyze Pioneer radiometric and telemetry data is part of a broader strategy that has the following objectives in sight \cite{stanford,problem_set_05}:
\begin{itemize}
\item To analyze the early mission data, with the aim of determining the direction of the anomaly in particular;
\item to study the physics of the planetary encounters;
\item to study the temporal evolution of $a_P$;
\item to compare the anomalous accelerations of the two Pioneers at similar heliocentric distances; and
\item to investigate the effect of on-board systematic forces on the spacecraft's motion and spin decay using telemetry data from the MDRs.
\end{itemize}
Below we shall discuss these objectives in more detail.

\subsection{Analysis of early data}
\begin{figure}[h!]
\centering
\psfig{figure=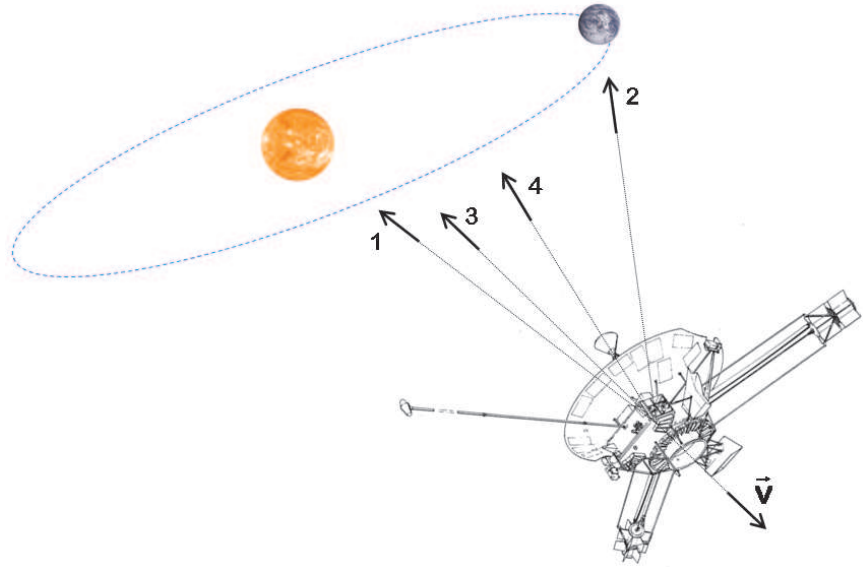,width=0.8\linewidth}
\protect\caption[The earlier part of the Pioneer 10 trajectory before Jupiter encounter. One can clearly see four possible different directions for the Pioneer anomaly.]{
The earlier part of the Pioneer 10 trajectory before Jupiter encounter, the part of the trajectory when antenna articulation was largest. One can clearly see four possible different directions for the Pioneer anomaly:
\begin{inparaenum}[i)]
\item toward the Sun,
\item toward the Earth,
\item along the velocity vector, and
\item along the spin axis.
\end{inparaenum}
Later in the mission the difference between these directions became very small.
}
\label{fig:directions}
\end{figure}

For the Pioneer spacecraft to maintain communications with the Earth using the narrow beam of their HGA, an Earth-pointing attitude was necessary. During the flight through the inner solar system, this required frequent attitude correction maneuvers to reorient the spin axis. At this time, the Sun--spacecraft--Earth angle was relatively large, as was the angle between these and the direction of motion (Fig.~\ref{fig:directions}).

Therefore, we expect that our analysis of the early data, from a period of time when the spacecraft were much closer to the Earth and the Sun, may help us to determine unambiguously whether the direction of the acceleration is
\begin{itemize}
\item pointing towards the Sun, indicating a force originating from the Sun;
\item pointing towards the Earth that would be due to an anomaly in the frequency standards;
\item along the direction of the velocity vector, indicating an inertial or drag force; or
\item along the spin axis direction that would indicate an on-board systematic force.
\end{itemize}

It should be pointed out that there are some obstacles along the way towards this goal. A difficult problem \cite{JPL2002,stanford,JPL2004} in deep space navigation is precise three-dimensional orbit determination. The line-of-sight component of a velocity is much more easily determined than motion in the orthogonal directions. Unfortunately there is no range observable for the Pioneer spacecraft, which complicates the analysis. Furthermore, earlier parts of the trajectory were dominated by solar radiation pressure and frequent attitude control maneuvers, which affect the accuracy of the orbit determination significantly. Nevertheless, there is hope that these difficulties can be overcome and the analysis will yield the true direction of the anomaly.

\subsection{Study of planetary encounters}

The early Pioneer 10 and 11 data (before 1987) were never analyzed in detail. However, for Pioneer 11, a small value for the anomaly was found during the Jupiter--Saturn cruise phase. Right at the time of the Saturn encounter, however, when the spacecraft passed into an hyperbolic escape orbit, there was a rapid increase in the anomaly, whereafter it settled into the canonical value.

We first plan to study the Saturn encounter for Pioneer 11. We plan to use the data for approximately 2 years surrounding this event. If successful, we should be able to find more information on the mechanism that led to the onset of the anomaly during the flyby.

\subsection{Temporal evolution of the anomaly}

If the anomalous acceleration is due to heat produced by the on-board nuclear fuel inventory, one expects a decrease in the anomaly's magnitude that will be correlated with the decay of the $^{238}$Pu fuel on board. The analysis of 11.5 years of data found no support for this mechanism. At the same time, although not very precise, the analysis performed in ref.~\citen{CBM2005} indicates a decay of the right magnitude in the solution for the anomaly. The much longer data span (30 years for Pioneer 10, 20 years for Pioneer 11) will help to better determine if there is any signature of an exponential decay of the on-board power source. At the same time, our analysis of the telemetry data will help us better profile the decay signature, by separately analyzing the thermal signature of the radioactive fuel source and of the electrical equipment on board.

\subsection{Analysis of the individual trajectories of the two Pioneers}

The much larger data set for Pioneers 10 and 11 makes it possible to study the properties of the individual solutions for both Pioneers. The data used in the previous analysis \cite{JPL2002,JPL1998,moriond} precluded comparison of the solutions for anomalous accelerations obtained with the data collected from the same heliocentric distances. The new data set will allow such an investigation. Previously, even though we had individual solutions from the two craft, the fact remained that $a_{P_{10}}$ and $a_{P_{11}}$ were obtained from data segments that were not only very different in length (11.5 and 3.75 years), but they were also taken from different heliocentric distances. In the new analysis, we will be able to examine the acceleration of the two Pioneers using comparable data sets, and we will also be able to utilize telemetry to understand the individual ``signatures'' of the two spacecraft and their potential impact on the acceleration.

\subsection{On-board systematic forces}

It is worth noting that no other mission in the past used telemetry data to improve its navigational capabilities. Our attempt to utilize Pioneer telemetry to refine the trajectory solutions is a novel technique.

\subsubsection{Propulsion}

Based on the information provided by the MDRs, one can develop a high accuracy thermal, electrical, and dynamical model of the Pioneer spacecraft. Such a model can be used to further improve our understanding of the anomalous acceleration and especially to study the contribution from the on-board thermal environment to the anomaly.

It is clear that a thermal model for the Pioneer spacecraft would have to account for all heat radiation produced by the spacecraft. In Sect.~\ref{sec:telemetry} we demonstrated how telemetry information can be used to accurately estimate the amount of heat produced by the spacecrafts' major components. The next step is to utilize this result along with information on the spacecrafts' design to estimate the amount of heat radiated in various directions.

We can also estimate the radio beam reaction force more precisely than previous studies. Instead of assuming that the power of the radio beam was always the nominal 8~W, we can use telemetry data to estimate the power of the radio beam as a function of time.
\begin{figure}[h!]
\begin{minipage}[b]{.5\linewidth}
\centering \psfig{file=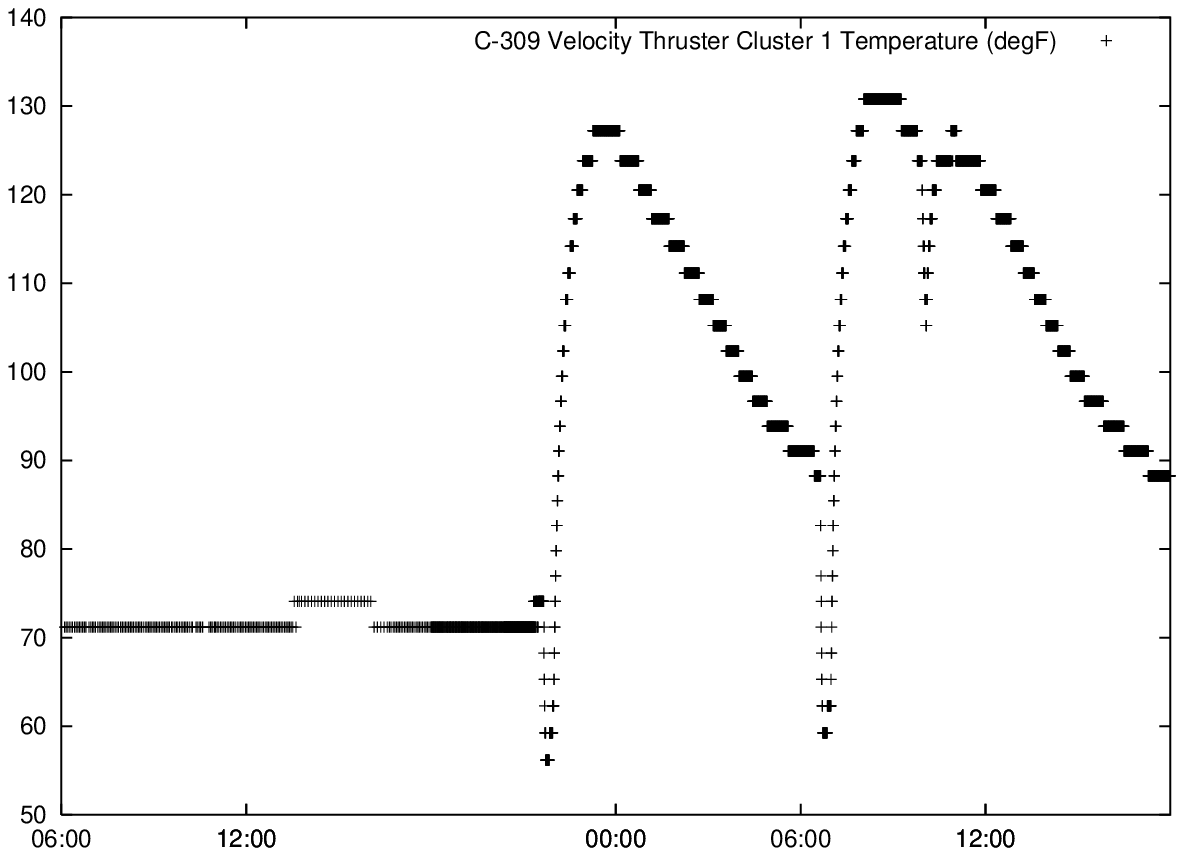, width=\linewidth}
\end{minipage}
\hskip 0.001\linewidth
\begin{minipage}[b]{.5\linewidth}
\centering \psfig{file=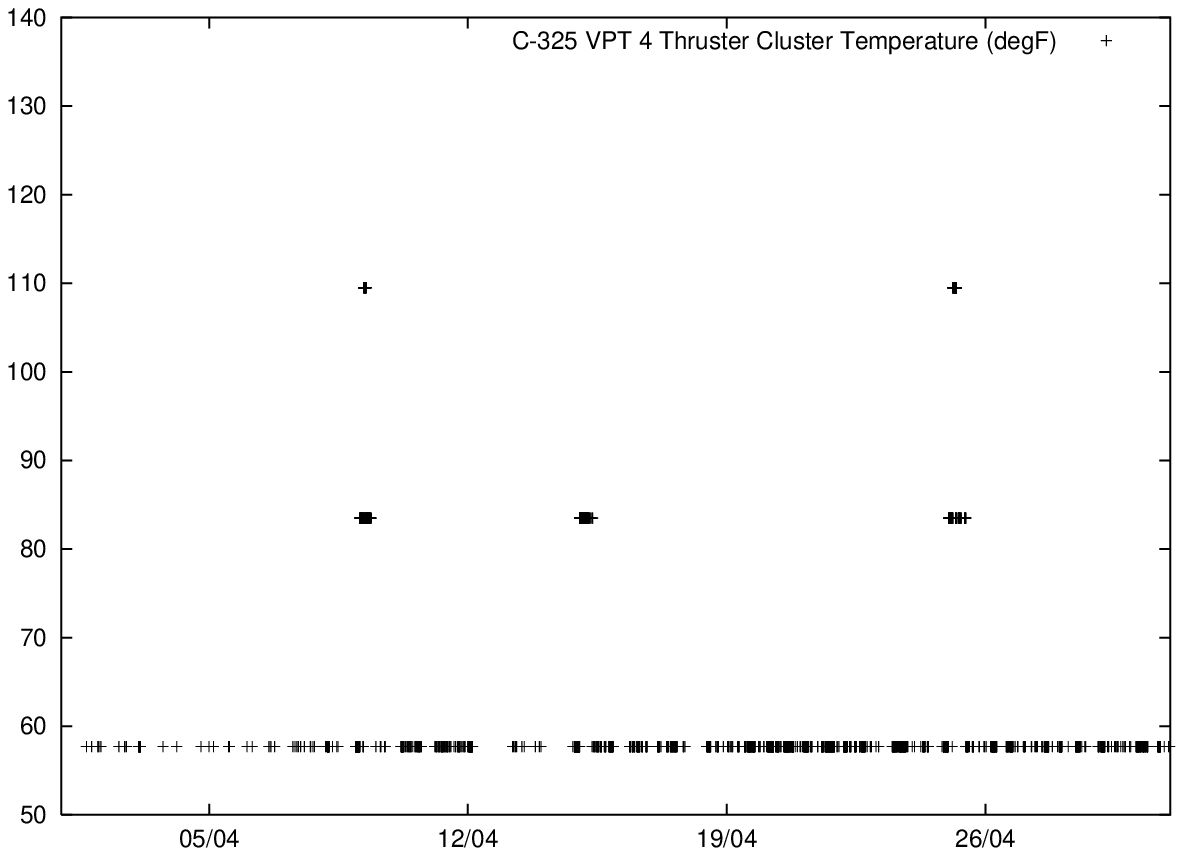, width=\linewidth}
\end{minipage}
\caption{Thruster temperature changes (in $^\circ$F) during maneuvers. During a major course correction maneuver on 19-20 April 1974, the temperature of Pioneer 11's velocity thruster cluster 1 (left) increased significantly. Less dramatic, but clearly visible are brief periods of elevated temperature in Pioneer 10's velocity and precession thruster 4 (right) during the month of April 1975, as the spacecraft executed minor attitude correction maneuvers.}
 \label{fig:P11_clusters}
\end{figure}

Telemetry also contains information relevant to maneuvers. For example, thruster cluster temperature readings (see Fig.~\ref{fig:P11_clusters}) can clearly indicate when a thruster was fired, which can be correlated with other sources of information on maneuvers. The rapid depletion of the fuel tank is also reflected in corresponding changes in fuel tank pressure and temperature.

\subsubsection{Spin rate behavior}

\begin{figure*}[ht]
\hskip -6pt
\begin{minipage}[b]{.5\linewidth}
\centering \psfig{file=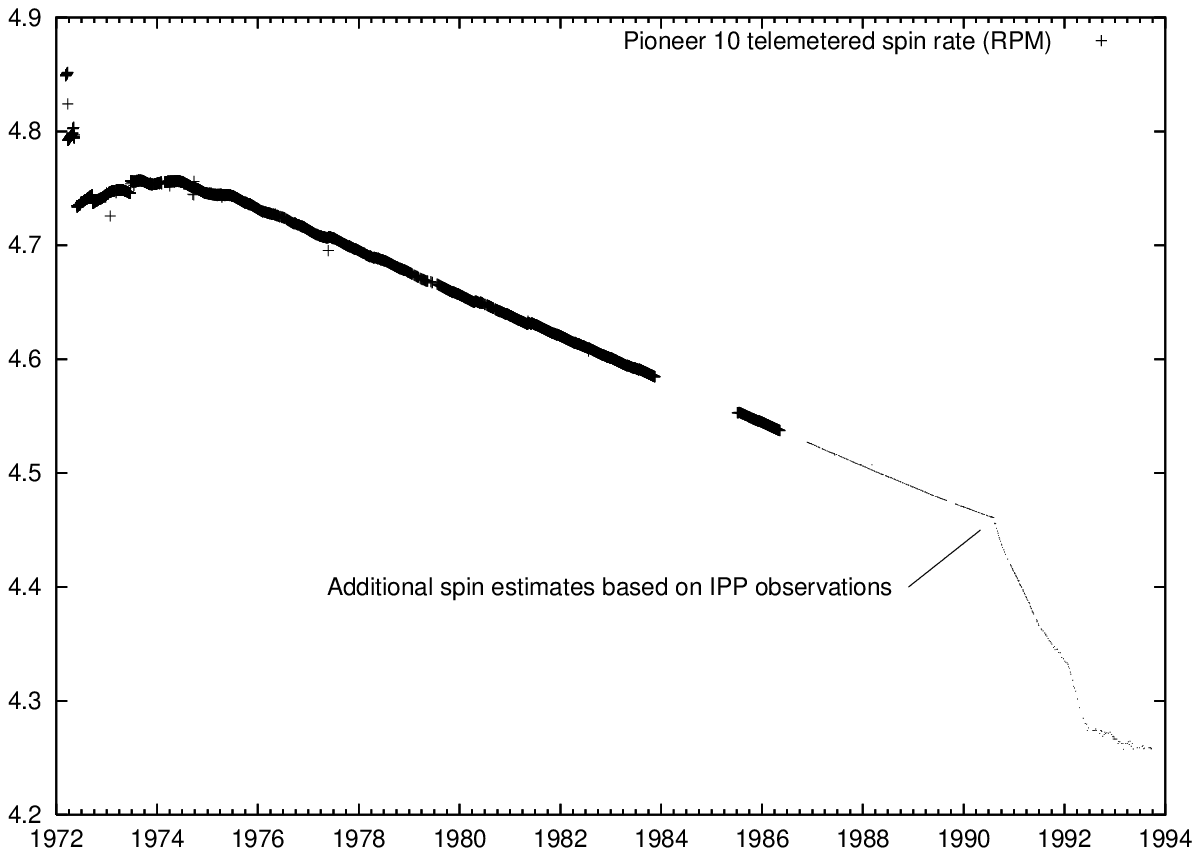, width=\linewidth}
\end{minipage}
\hskip 0.001\linewidth
\begin{minipage}[b]{.5\linewidth}
\centering \psfig{file=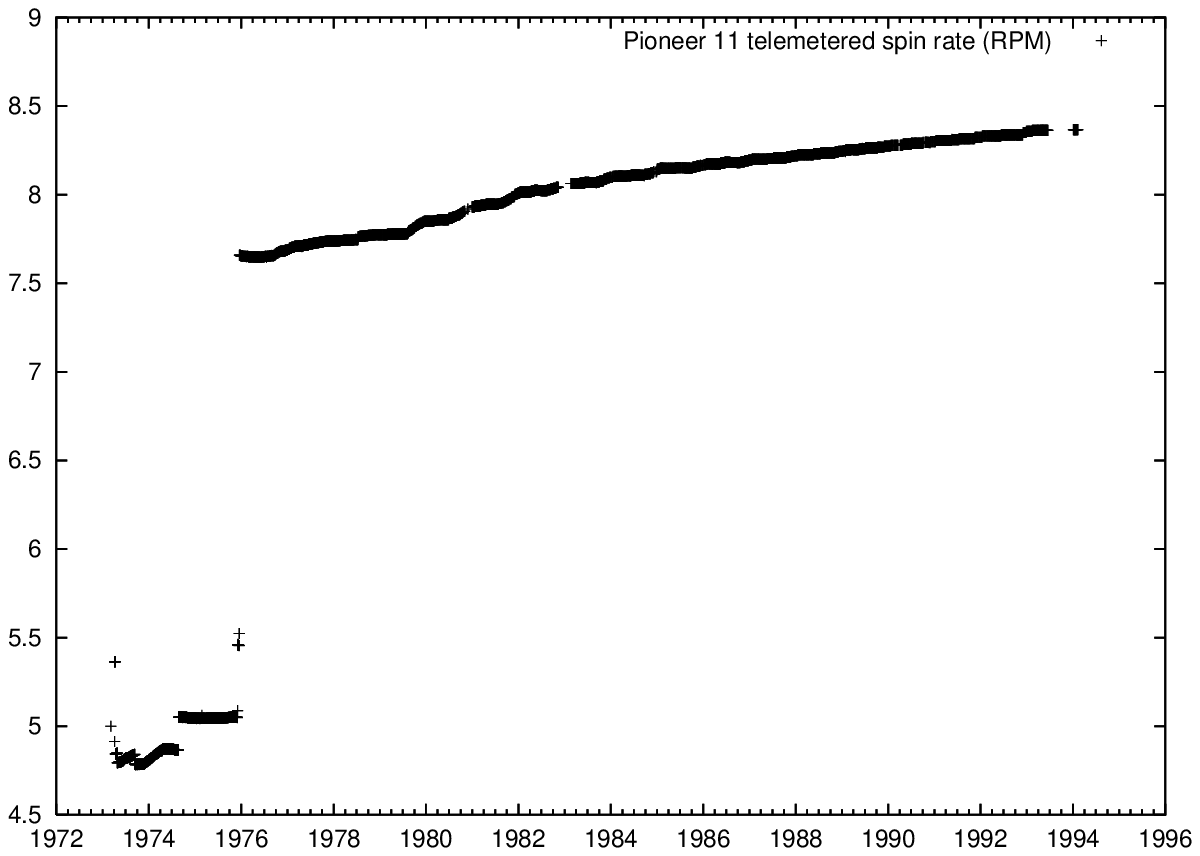, width=\linewidth}
\end{minipage}
\caption{On-board spin rate measurements (in RPM) for Pioneer 10 (left) and Pioneer 11 (right). The sun sensor used on Pioneer 10 for spin determination was temporarily disabled between November 1983 and July 1985, and was turned off in May 1986, resulting in a `frozen' value being telemetered that no longer reflected the actual spin rate of the spacecraft. Continuing spot measurements of the spin rate were made using the Imaging Photo-Polarimeter (IPP) until 1993. The anomalous increase in Pioneer 11's spin rate early in the mission was due to a failed spin thruster. Continuing increases in the spin rate were due to maneuvers; when the spacecraft was undisturbed, its spin rate slowly decreased, as seen in Figure~\ref{fig:spinzoom}.}
\label{fig:spin}
\vskip -5pt
\end{figure*}

The Pioneer 10 and 11 spacecraft were spin-stabilized, with a nominal spin rate of 4.8~rpm. The actual spin rate of Pioneer 10 remained very close to this value throughout its mission. The spin rate of Pioneer 11 was much higher, around 8~rpm, due to a spin thruster failure early in the mission. The spin rate of both spacecraft changed over time (Fig.~\ref{fig:spin}).

As the spacecraft's transmission is a circularly polarized radio beam, the spin of the spacecraft influences the transmission frequency \cite{JPL2002,Anderson_Mashhoon-2003}. Every revolution of the spacecraft adds to (or subtracts from) the transmission one cycle of phase. Since the spacecrafts' spin rate were not constant, this introduces a variable bias in the Doppler analysis.

The Pioneer spacecraft utilized a redundant pair of sun sensors and a single star sensor to accurately determine their spin rate. The sensors generated a pulse at every spacecraft revolution. The time between pulses was measured on board with an accuracy of $1/8192$~s, and this reading was regularly telemetered to the ground.

Unfortunately, the star sensor on board Pioneer 10 became inoperative after the encounter with Jupiter \cite{AIAA870502}. Subsequently, a sun sensor was used for spin rate determination. However, the sun sensor fails when the spacecraft is too far away from the Sun ($\gtrsim33$~AU), or when the angle between the direction of the Sun and the spacecraft spin axis becomes small. The latter condition occurred after an orientation maneuver in November 1983, causing the spin rate to ``freeze'' at the last reported value until July 1985, when the sun sensor was reenabled. The sun sensor was turned off for the final time in May 1986, and from that point onward, the spin rate telemetry continued to report the last measured reading, no longer reflecting the actual spin rate of the spacecraft.

Afterwards, one of the spacecraft's instruments (the Imaging Photo-Polarimeter) was used occasionally to determine the position of a known star relative to the spacecraft's orientation. Changes in this reading were used to calculate the spin rate on the ground. The actual spin rate continued to decrease, as reported for instance in ref.~\citen{JPL2002}.\footnote{The nature of this nearly linear decrease continues to be a mystery and will be further investigated during the upcoming analysis of the entire sets of both Doppler and telemetry data received from Pioneers 10 and 11.}

For Pioneer 11, the star sensor remained operational throughout the mission and good spin rate readings were telemetered.

\begin{figure*}
\hskip -6pt
\begin{minipage}[b]{.5\linewidth}
\centering \psfig{file=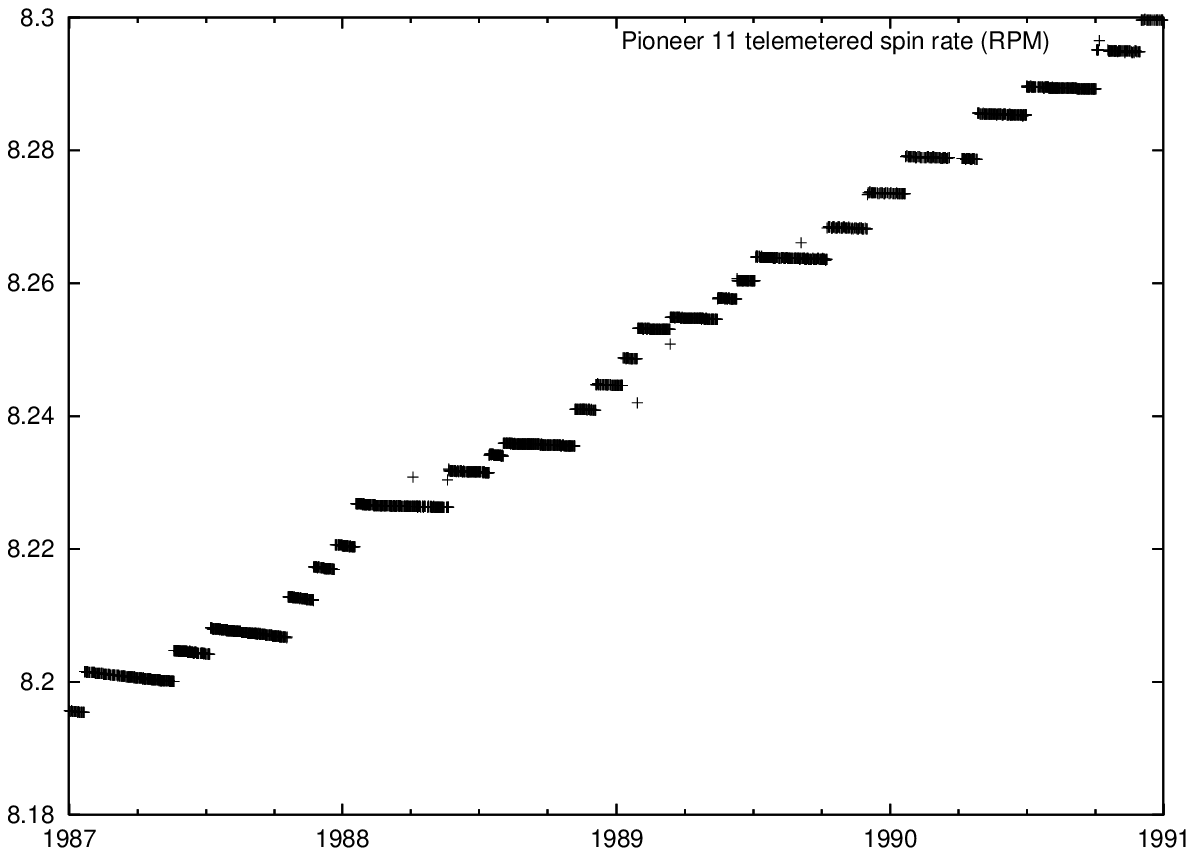, width=\linewidth}
\end{minipage}
\hskip 0.001\linewidth
\begin{minipage}[b]{.5\linewidth}
\centering \psfig{file=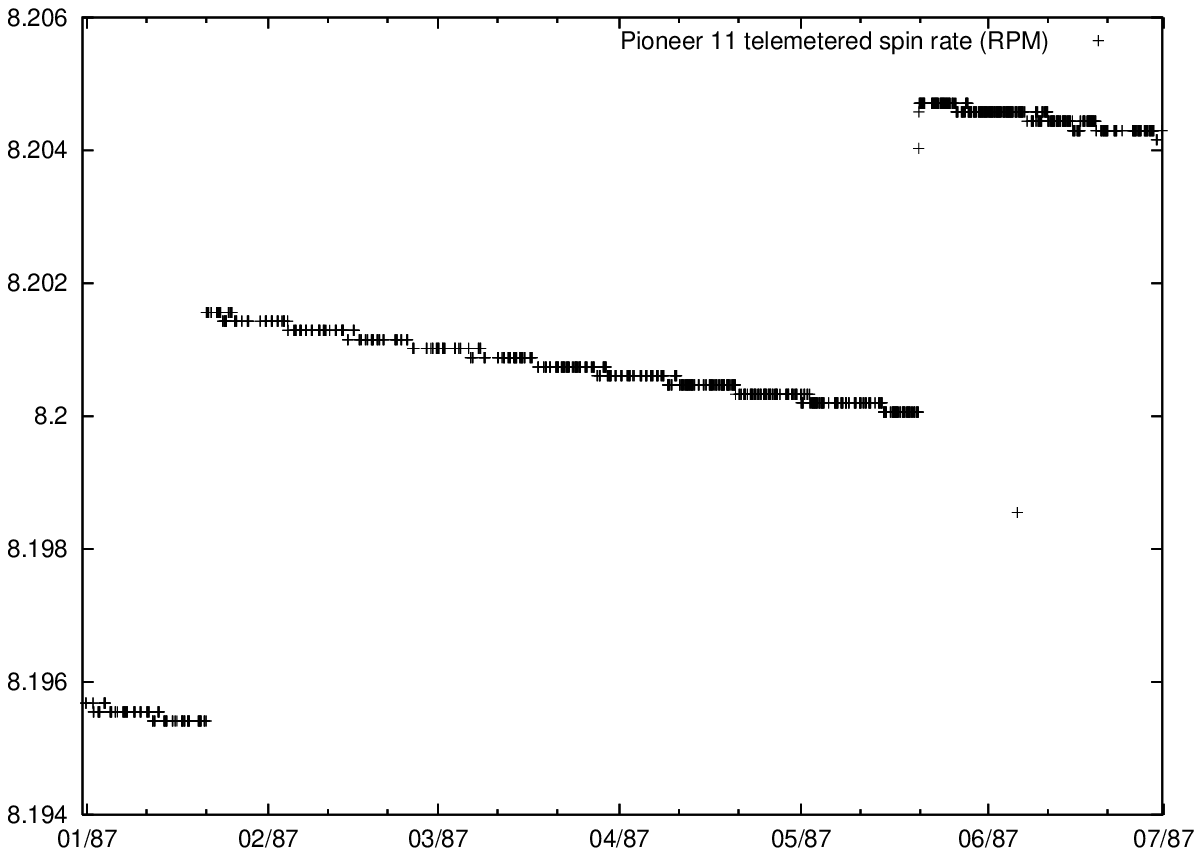, width=\linewidth}
\end{minipage}
\caption{Zoomed plots of the spin rate of Pioneer 11. On the left, the interval examined in ref.~\citen{JPL2002} is shown; maneuvers are clearly visible, resulting in discrete jumps in the spin rate. The figure on the right focuses on the first half of 1987; the decrease in the spin rate when the spacecraft was undisturbed is clearly evident.}
\label{fig:spinzoom}
\vskip -5pt
\end{figure*}

\section{Implications for other missions}

The methods we develop to evaluate the magnitude and behavior of the anomalous acceleration of the Pioneers 10 and 11, and to estimate the contribution of on-board systematic forces to the anomaly, may be directly applicable to other missions.

The first mission that comes to mind is NASA's recently launched New Horizons spacecraft to Pluto. Although there are many differences in their design, New Horizons and Pioneer 10 and 11 share some important features. Both spacecraft are spin-stabilized, utilizing a minimum number of thruster firings to maintain attitude. Both spacecraft are dominated in appearance by an Earth-pointing HGA. Both spacecraft use RTGs to generate electricity. Both spacecraft utilize a louver system to vent excess heat from the spacecraft body.

Among the differences is the fact that the single RTG on New Horizons is much closer to the body of the spacecraft and, therefore, more of its thermal radiation will be reflected off the back of the HGA, increasing the anisotropy of the thermal radiation pattern of this spacecraft.  The thermal output of the single RTG on New Horizons is nearly twice the combined thermal output of the four SNAP-19 RTGs that powered one Pioneer spacecraft.  The New Horizons RTG generates more electricity, which ultimately is converted to heat inside the spacecraft and then preferentially radiated away in a direction opposite the HGA.  For these reasons, the contribution of on-board heat to the unmodeled acceleration of New Horizons will be larger and will somewhat degrade the navigational accuracy for this mission. In fact, it  was recently realized that a number of on-board sources of systematic effects, most notably the radiant heat emitted by the RTGs, will limit New Horizons' acceleration sensitivity at the level of $\sim 4\times10^{-9}$ m/s$^2$, affecting some mission objectives.

Fortunately, relying on our recent experience of working with the Pioneer spacecraft, we can demonstrate that, by using the on-board telemetry, New Horizons can still accomplish all its radio-science objectives. To achieve this goal, one needs to build a high accuracy thermal-electric-dynamical model for the New Horizons spacecraft.  Because of the availability of the relevant data and for the purposes of proving the feasibility of the approach, we first embarked on the development of such a model for the Pioneer spacecraft, as discussed in ref.~\citen{new-data-2005}. However, once the appropriate data (radiometric and telemetry) is available from New Horizons, a similar model will be developed for this mission to allow a ``real-time'' calibration of navigational data with respect to the set of on-board systematics. As a quantitative objective, we hope to improve acceleration sensitivity on this spacecraft to the level below $2\times10^{-10}$~m/s$^2$. In return, this will lead to an improvement of the spacecraft's navigation accuracy, increasing the mission's science return. A detailed analysis of the impact of this effect on the trajectory of the New Horizons spacecraft has been initiated and will be reported elsewhere.

Therefore, the study of the Pioneer anomaly already demonstrates the value of telemetry in improving navigational accuracy of future spacecraft by calibrating for systematic effects generated on board. Our ability to reconstruct a timeline for the Pioneer spacecraft is wholly dependent on the availability of raw telemetry and radiometric data, which strongly advocates for the preservation of the {\em entire} raw electronic record of all space missions.

We emphasize the fact that no other mission in the past used telemetry data to improve its navigational capabilities.  Our approach will result in a new method to increase the navigational accuracy of deep space probes (an alternative to VLBI, accelerometers, drag-free technologies); an improvement by more than an order of magnitude appears possible at no additional cost. Furthermore, there is already a recognition that understanding of the on-board systematics may lead to improved attitude control systems for low disturbance spacecraft for the needs of fundamental physics---the most significant outcome of our current work on the Pioneer anomaly.

\section{Conclusions}

By 2006, the existence of the Pioneer anomaly is no longer in doubt. A steadily growing part of the community has concluded that the anomaly should be subject to further investigation and interpretation. A very important first step in this direction is to improve our knowledge of the anomaly by examining the entire mission record to the extent possible.

This entails, on the one hand, an analysis of all available radiometric data, to characterize the anomalous acceleration beyond the periods that were examined in previous studies. Telemetry, on the other hand, enables us to reconstruct a thermal, electrical, and propulsion system profile of the two spacecraft. Soon, we expect to be able to accurately estimate any effects on the motion of the spacecraft due to on-board systematic acceleration sources, expressed as a function of telemetry readings. This provides a new and unique way to refine orbital predictions and will also lead to an unambiguous determination of the origin of the Pioneer anomaly.

We would like to thank Larry Kellogg for his extraordinary effort to preserve the Pioneers' raw telemetry data records. We also express our deep appreciation of the insights in understanding of the retrieved  radiometric Doppler data obtained from our colleagues at JPL, especially Eunice L. Lau, and Kyong J. Lee.   As many times before, David Lozier of NASA/Ames was particularly helpful.

The work of SGT was carried out at the Jet Propulsion Laboratory, California Institute of Technology, under a contract with the National Aeronautics and Space Administration.

\end{document}